\documentstyle[11pt,aaspp]{article}

\newcommand{\eps} {\varepsilon}
\newcommand{\fch} {{f_R}}
\newcommand{\muc} {{\mu_R}}
\newcommand{\lone} {\lambda_1}
\newcommand{\ltwo} {\lambda_2}
\newcommand{\ce} {{\cal E}}
\newcommand{\cc} {{\cal C}}
\newcommand{\Ibar}{{\hbox{\rlap{\hbox{\raise.25ex\hbox{-}}}
   I\llap{\hbox{\raise.25ex\hbox{-}}}}}}
\newcommand{\go}{\gtrsim}
\newcommand{\lo}{\lesssim}

\begin{document}

\title{
HYDRODYNAMIC INSTABILITY AND COALESCENCE\\ OF BINARY NEUTRON STARS
}

\medskip

\author{Dong Lai\altaffilmark{1}}
\affil{Center for Radiophysics and Space Research, Cornell University, Ithaca,
NY 14853}
\author{Frederic A. Rasio\altaffilmark{2}}
\affil{Institute for Advanced Study, Olden Lane, Princeton, NJ 08540}
\and
\author{Stuart L. Shapiro\altaffilmark{3}}
\affil{Center for Radiophysics and Space Research, Cornell University,
Ithaca, NY 14853}
\altaffiltext{1}{Department of Physics, Cornell University.}
\altaffiltext{2}{Hubble Fellow.}
\altaffiltext{3}{Department of Astronomy and Department of Physics, Cornell
University.}

\medskip

\begin{abstract}
We study the importance of hydrodynamic effects on the evolution
of coalescing binary neutron stars.
Using an approximate energy functional constructed from equilibrium solutions
for polytropic binary configurations, we incorporate hydrodynamic effects
into the calculation of the orbital decay driven by
gravitational wave emission.
In particular, we follow the transition between the quasi-static, secular decay
of the orbit at large separation and the rapid dynamical evolution
of configurations approaching contact. We show that a purely
Newtonian hydrodynamic instability can significantly
accelerate the coalescence at small separation.
Such an instability occurs in all close
binary configurations containing sufficiently incompressible stars.
Calculations are performed for various neutron star masses, radii, and
spins. The influence of the stiffness of the equation of state is
also explored by varying the effective polytropic index.
Typically, we find that the radial infall velocity just prior to contact
is about $10\%$ of the tangential orbital velocity. Once the stability
limit is reached, the final evolution only takes another orbit.
Post-Newtonian effects can move the stability limit to a larger
binary separation, and may induce an even larger radial velocity.
We also consider the possibility of mass transfer from one neutron star to
the other. We show that stable mass transfer is impossible except when the
mass of one of the components is very small ($M\lo 0.4 M_{\odot}$)
and the viscosity is high enough to maintain corotation.
Otherwise, either the two stars come into contact or the
dynamical instability sets in before a Roche limit can
be reached.
\end{abstract}

\keywords{hydrodynamics -- instabilities -- stars: neutron -- stars: rotation
--
 stars: binaries: close -- radiation mechanisms: gravitational}

\section{INTRODUCTION}

Coalescing neutron star binaries have long been recognized as a
most promising source of gravitational radiation that
could be detected by the new generation of laser interferometers such as LIGO
(Clark 1979; Thorne 1987; Abramovici et al.~1992; Cutler et al.~1992).
Statistical arguments based on the observed local
population of binary pulsars and type Ib supernovae lead to an estimate
of the rate of neutron star binary coalescence in the Universe of
about $10^{-7}\,$yr$^{-1}\,$Mpc$^{-3}$
(Narayan, Piran \& Shemi 1991; Phinney 1991).
Finn \& Chernoff (1993) estimate that an advanced LIGO detector could observe
about 70 events per year. In addition to providing a major confirmation of
Einstein's theory of general relativity, the detection of gravitational
waves from coalescing binaries at cosmological distances could provide
the first accurate measurement of the Universe's Hubble constant
and mean density (Schutz 1986; Cutler et al.~1992).
Coalescing neutron stars may also produce intense bursts of
neutrinos and high energy photons, and could  be the source of
extra-galactic gamma-ray bursts (Eichler et al.~1988; Paczy\'nski 1986, 1991).

Recent calculations of the gravitational radiation waveforms
from coalescing neutron star binaries
have focused on the signal emitted during the last few thousand orbits,
as the frequency sweeps upward from about 10$\,$Hz to 1000$\,$Hz.
The waveforms in this regime can be calculated fairly accurately by
performing high-order post-Newtonian expansions of the equations of
motion for two {\em point\/} masses
(Lincoln \& Will 1990; Junker \& Schafer 1992).
Accuracy is essential here since the observed signal will be matched against
theoretical templates. Since the templates must cover $>10^3$ orbits,
a fractional error as small as $10^{-3}$ can prevent detection.
When, at the end of the inspiral,  the binary separation becomes comparable
to the stellar radii, hydrodynamic effects become important and the character
of the waveforms will be different.
Special purpose narrow-band detectors that can sweep up frequency in real
time will be used to try to catch the final few cycles of gravitational
radiation (Meers 1988; Strain \& Meers 1991).
In this terminal phase of the coalescence,
the waveforms should contain information not just about the
effects of relativity,
but also about the internal structure of neutron stars. Since the masses and
spins of the two stars, as well as the orbital parameters can be determined
very accurately from the lower-frequency inspiral waveform,
a simple determination of the maximum frequency $f_{max}$ reached by
the signal should provide a measurement of
the neutron star radii and place severe constraints on nuclear equations
of state (Cutler et al.~1993). Such a measurement requires theoretical
knowledge about all relevant hydrodynamic processes.

Two regimes can be distinguished in which different hydrodynamic processes
take place. The first regime corresponds to the 10 or so orbits preceding
the moment when the surfaces of the two stars first come into contact.
In this regime, the two stars are still approaching each other in
a quasi-static manner, but the tidal effects are very large.
The second regime corresponds to the subsequent merging of the two stars
into a single object. This involves very large departures from
hydrostatic equilibrium, including mass shedding and shocks,
and can be studied only by means of three-dimensional,
fully hydrodynamic computations (Nakamura \& Oohara 1991 and references
therein; Rasio \& Shapiro 1992)
By contrast, in the first regime, the evolution of the system can be described
fairly accurately by a sequence of near-equilibrium fluid configurations.
We adopt such a description in this paper.
Since neutron stars are not very compressible, the equilibrium configurations
are not very centrally condensed and the usual Roche model for close binaries
(e.g., Kopal 1959) does not apply. Instead,
the hydrostatic equilibrium equation must be
solved in three dimensions for the structure of the system.

In two recent papers (Lai, Rasio \& Shapiro 1993b,~c,
hereafter LRS1 and LSR2), we
have used an approximate energy variational method to study analytically the
hydrostatic equilibrium
and stability properties of Newtonian binary systems obeying polytropic
equations of state.
In particular, we have constructed the compressible generalizations
of all the classical solutions for binaries containing
incompressible  ellipsoidal components (Chandrasekhar 1969).
The so-called {\em Darwin-Riemann configurations\/}
are  of special importance for modeling  neutron star binaries.
They are generalizations of the Roche-Riemann
configurations originally introduced by
Aizenman (1968) to describe an incompressible star in
circular orbit around a point mass\footnote{A Roche-Riemann
binary consists of a star and point mass, while a Darwin-Riemann
binary consists of two finite-size stars.}.
Kochanek (1992) first pointed out the importance of these
configurations for modeling neutron star binaries.
In Darwin-Riemann binaries of the type considered here,
fluid motions with uniform vorticity parallel to the rotation axis
are present in the corotating frame of the binary system.
These configurations represent a simple model of a close binary
containing stars whose spins are not necessarily synchronized
with the orbital motion.
Indeed, it has been argued recently that the synchronization time
of neutron star binaries may remain very long compared to their
orbital decay timescale (Bildsten \& Cutler 1992; Kochanek 1992).
In the limit where the viscosity is negligible (leading to an
infinite synchronization time), the fluid circulation is strictly
conserved and an asynchronous configuration of the Darwin-Riemann type is
expected (Miller 1974; Kochanek 1992).
The value of the circulation is determined by the
initial spins of the two neutron stars at large
binary separation. If the stars spin uniformly at large separation,
the vorticity will remain uniform. Similarly, if the spins are initially
aligned with the rotation axis of the binary (a most likely configuration), the
vorticity vector will remain aligned as well. Detailed calculations
of the properties of Darwin-Riemann binaries are presented in LRS1
(for two identical components) and LRS2 (for nonidentical components).

An important result found in LRS1 is that {\em Darwin-Riemann
configurations containing a sufficiently incompressible fluid can
become hydrodynamically unstable\/}. Close binaries containing
neutron stars with stiff equations of state ($\Gamma\ga2$)
should therefore be particularly susceptible to these instabilities.
As the dynamical stability limit is approached, the secular orbital
decay driven by gravitational wave emission can be dramatically accelerated.
It is this approach towards instability that we intend to study here.
When the stability limit is reached, the two stars plunge almost radially
toward each other, and merge together
into a single object after just a few orbits.
The complicated three-dimensional hydrodynamics characterizing this
very short terminal phase of the evolution can only be studied
with large-scale computer simulations (Nakamura \& Oohara 1991;
Rasio \& Shapiro 1992). In contrast, the long
transition between the very slow secular decay of the orbit at large separation
and the more rapid inspiral as the stability limit is approached is better
explored using quasi-analytic methods. The analytic results should also prove
helpful in the construction of better initial
conditions for the numerical simulations.
Indeed, nonsynchronized initial conditions are particularly difficult to deal
with numerically (for recent attempts, see Shibata et al.~1993;
Davies et al.~1993).

In a previous study (Lai, Rasio, \& Shapiro 1993a, hereafter Paper I),
we modeled the hydrodynamic and tidal effects associated with
finite-size, spinning components by modifying
the orbital evolution equations for two point masses
in a simple, heuristic way.
In this paper, we improve our treatment by introducing the full binary
equilibrium solutions for polytropes calculated in LRS1.
Specifically, we model a neutron star binary as a Darwin-Riemann
configuration containing two triaxial polytropes.
We calculate the orbital evolution of the
system as the stability limit is approached and we study how the
radial infall evolves from secular to dynamical.
Because of the simplicity of our method, we can perform a large
number of calculations, systematically surveying all relevant parameters
such as the stellar masses, radii, and spins. We also vary the
polytropic index of the fluid to mimic the mean stiffness of the
nuclear equation of state. All of these parameters can be chosen
separately for each of the two companion stars. However,
we focus our attention on binaries containing two identical neutron stars.
Indeed, all the measured masses of neutron stars in our Galaxy appear to be
narrowly clustered around $1.4M_{\sun}$ (Thorsett et al.~1993).
An important simplifying assumption which remains in the present
treatment is that at any separation the {\em internal\/} degrees of
freedom of the two stars (surface deformations and central densities)
are calculated from the values in circular
equilibrium while we follow the orbital
decay. This approximation is reasonable whenever the orbital timescale
remains shorter than the radial infall timescale, but longer than the internal
dynamical timescale of each star.
In a future paper, we will derive and solve the generalized
evolution equations for a binary configuration in which all degrees of freedom
are allowed to vary dynamically.

In \S 2 we present our Darwin-Riemann equilibrium model for binaries
containing two identical stars.
We also mention briefly Darwin-Riemann models for two stars with
 unequal masses, as well as Roche-Riemann models of
 neutron-star-black-hole binaries.
The orbital evolution of these models is calculated in \S 3 and \S 4.
General relativistic corrections to the orbital motion
 are discussed in \S 5. In \S 6, we
examine the different types of terminal evolutions that are
possible when the two stars have different masses. In particular,
we address the possibility that stable mass transfer from
one neutron star to the other may occur (as first suggested
by Clark \& Eardley 1977).

\clearpage

\section{DARWIN-RIEMANN EQUILIBRIUM MODELS FOR CLOSE BINARIES}

\subsection{Basic Equations}

In this section, we derive equilibrium equations for compressible
Darwin-Riemann
binary configurations containing two identical polytropes.
We use an energy variational method together with
an ellipsoidal approximation to obtain those equations.
Under the combined effects of centrifugal and
tidal forces, each polytrope assumes a nonspherical equilibrium
configuration which we approximate by a triaxial ellipsoid.
As in the classical Darwin problem,
we assume that the two ellipsoidal figures
corotate with the orbital angular velocity (i.e., we consider only
stationary tides), but, as in the classical Riemann ellipsoids of type S,
we allow for internal fluid motions with uniform
vorticity parallel to the rotation axis.
More details about the method and a number of applications to other types
of binary and isolated rotating configurations can be found in LRS1 and LRS2.

Consider a binary system containing two identical polytropes of mass $M$
and polytropic index $n$, in circular orbit about each other.
The density and pressure are related by
\begin{equation}
P=K\rho^{\Gamma},
\end{equation}
where the adiabatic exponent $\Gamma=1+1/n$. The constant $K$  measures the
specific entropy and is the same for both stars.
We denote the central density by $\rho_c$ and
the binary separation by $r$. We make the assumption that
the surfaces of constant density within each star
can be approximated by {\em self-similar ellipsoids\/}.
The geometry of the configuration is then completely specified by the three
principal axes of the outer surfaces, $a_1$, $a_2$, and $a_3$,
with $a_1$ measured along the axis of the binary,
$a_2$ in the direction of the orbital motion, and $a_3$ along
the rotation axis. In addition, we assume that the density profile $\rho(m)$,
where $m$ is the mass interior to an isodensity surface,
is identical to that of a spherical polytrope of same volume and entropy.

Under these assumptions, the total internal energy in the
system is given by
\begin{equation}
U=2 \int n \frac{p}{\rho}\, dm = 2 k_1K\rho_c^{1/n}M,    \label{Udef}
\end{equation}
and the total self-gravitational potential energy (setting $G=1$) is
\begin{equation}
W=-2\times \left(\frac{3}{5-n}\right)
\frac{M^2}{R}f=-2 k_2M^{5/3}\rho_c^{1/3} f.
\label{Wdef}
\end{equation}
In this expression,  we have introduced the mean radius
$R\equiv (a_1a_2a_3)^{1/3}$ and the dimensionless quantity
\begin{equation}
f\equiv \frac{A_1a_1^2+A_2a_2^2+A_3a_3^2}{2 R^2},      \label{fdef}
\end{equation}
such that $f=1$ for spherical stars.
The dimensionless coefficients $A_i$ are defined as in
Chandrasekhar (1969, \S17); they depend only on the axis  ratios
$a_3/a_1$ and $a_3/a_2$.
The coefficients $k_1$ and $k_2$ are dimensionless polytropic
structure constants which depend only on $n$,
\begin{equation}
k_1\equiv\frac{n(n+1)}{5-n}\,\xi_1|{\theta'}_1|,~~~~~~~~
k_2\equiv\frac{3}{5-n}\,\left(\frac{4 \pi |{\theta'}_1|}{\xi_1}\right)^{1/3},
    \label{kdef}
\end{equation}
where $\theta$ and $\xi$ are the usual Lane-Emden variables for a polytrope
(see, e.g., Chandrasekhar 1939).

The gravitational interaction energy $W_i$ between the two stars
is given, to quadrupole order, by
\begin{equation}
W_i=-\frac{M^2}{r}-\frac{M}{r^3}(2I_{11}-I_{22}-I_{33}),    \label{Widef}
\end{equation}
where
\begin{equation}
I_{ij}=\kappa_n \frac{M a_i^2}{5}\delta_{ij}~~~~~~~~~~~{\rm (no~sum)}
\label{Iijdef}
\end{equation}
and $\kappa_n$ is a constant depending only on $n$,
\begin{equation}
\kappa_n\equiv\frac{5}{3\xi_1^4|{\theta'}_1|}\,\int_0^{\xi_1}\theta^n\xi^4
 \,d\xi,          \label{kndef}
\end{equation}
so that $\kappa_n=1$ for $n=0$. Values of $k_1,~k_2$ and $\kappa_n$ for
different $n$ are given in Table 1 of LRS1.

Now turn to the kinetic energy.
For a synchronized binary, the fluid is simply in uniform rotation at
the orbital frequency $\Omega{\bf e}_3$ perpendicular to the
orbital plane. Here, however, we allow for
an additional internal motion of the fluid inside each star, assumed to have
uniform vorticity $\zeta{\bf e}_3$ as measured in the corotating frame
of the binary,
\begin{equation}
\zeta\equiv(\nabla\times {\bf u})\cdot {\bf e}_3
=-\frac{a_1^2+a_2^2}{a_1a_2}\Lambda,    \label{zdef}
\end{equation}
where
\begin{equation}
{\bf u}=\frac{a_1}{a_2}\Lambda x_2 \,{\bf e}_1
-\frac{a_2}{a_1}\Lambda x_1 \,{\bf e}_2      \label{Ldef}
\end{equation}
is the fluid velocity in the corotating frame. Here the origin of
the coordinates $x_i$ is at the center of mass of the star. The
quantity $\Lambda$ is the angular frequency of the internal fluid motions.
Note that the velocity field (\ref{Ldef}) is everywhere tangent to the
ellipsoidal surfaces of constant density.
For a synchronized binary system, we have ${\bf u}=\zeta=\Lambda=0$.
The velocity field in the inertial frame relative to the center
of mass of the star is given by
\begin{equation}
{\bf u}^{(0)}={\bf u}+{\bf \Omega}\times{\bf x}.   \label{u0def}
\end{equation}
The vorticity in the inertial frame is
\begin{equation}
{\zeta}^{(0)}\equiv(\nabla\times{\bf u}^{(0)})\cdot {\bf e}_3
=(2+\fch)\Omega,
\end{equation}
where we have defined the ratio
\begin{equation}
\fch\equiv{\zeta\over\Omega}.
\end{equation}
It is straightforward to calculate the total kinetic energy
corresponding to this velocity field. We find
\begin{equation}
T = T_s+T_o=2 \left[\frac{1}{2}I(\Lambda^2+\Omega^2)
-\frac{2}{5}\kappa_nMa_1a_2\Lambda\Omega \right]
+\frac{1}{2} \mu r^2 \Omega^2,    \label{Tdef}
\end{equation}
where $\mu=M/2$ is the reduced mass and
$I=I_{11}+I_{22}=\kappa_n M(a_1^2+a_2^2)/5$ is the
moment of inertia of each star. We have also defined the orbital
kinetic energy $T_o\equiv(\mu/2)r^2\Omega^2$ and the ``spin''
kinetic energy $T_s\equiv T-T_o$.
The total angular momentum $J$ can be written similarly as
\begin{equation}
J = 2 (I\Omega -\frac{2}{5}\kappa_nMa_1a_2\Lambda)
+ \mu r^2 \Omega. \label{Jdef}
\end{equation}
Another important conserved quantity is the fluid circulation $C$ along
the equators of the two star. Following LRS1 we write
\begin{equation}
\cc\equiv \biggl (-{1\over5\pi}\kappa_nM\biggr)C
=2\biggl (-{1\over5\pi}\kappa_nM\biggr)\pi a_1a_2 {\zeta}^{(0)}
=2 (I\Lambda-\frac{2}{5}\kappa_nMa_1a_2\Omega).      \label{Cdef}
\end{equation}
Note that $\cc$ has dimensions of an angular momentum but is proportional
to the conserved circulation. We usually refer to $\cc$ itself as the
circulation. Using equations~(\ref{Jdef}) and~(\ref{Cdef}),
the kinetic energy can be rewritten as
\begin{equation}
T=T_s+T_o
=T_{+}+T_{-}+\frac{1}{2}\mu r^2 \Omega^2,    \label{T2def}
\end{equation}
where
\begin{equation}
T_{\pm}=\frac{1}{4}I_{\pm}(\Omega \pm \Lambda)^2
=\frac{1}{4I_{\pm}}(J \pm \cc -\mu r^2 \Omega)^2,
\end{equation}
and
\begin{equation}
I_{\pm}=\frac{2}{5} \kappa_n M(a_1 \mp a_2)^2=2I_s/h_{\pm},
\end{equation}
with $I_s=2\kappa_nMR^2/5$ and $h_{\pm}=2R^2/(a_1\mp a_2)^2$.

The total energy of the system (not necessarily in equilibrium) can now be
written simply as the sum of expressions (\ref{Udef}), (\ref{Wdef}),
(\ref{Widef}), and~(\ref{T2def}),
\begin{equation}
E(\rho_c,\lambda_1,\lambda_2,r;\,M,J,\cc)=U+W+W_i+T,    \label{Edef}
\end{equation}
where we have chosen as independent variables the central density
$\rho_c$, the binary separation $r$, and the two {\em oblateness parameters\/}
$\lambda_1 \equiv (a_3/a_1)^{2/3}$ and $\lambda_2 \equiv (a_3/a_2)^{2/3}$.
The equilibrium structure of the binary can be determined
from the four conditions
\begin{equation}
\frac{\partial E}{\partial r}=\frac{\partial E}{\partial\rho_c}
=\frac{\partial E}{\partial\lambda_1}
=\frac{\partial E}{\partial\lambda_2}=0.     \label{equil}
\end{equation}
The first condition, $\partial E/\partial r=0$, gives the
equilibrium relation between $\Omega^2$ and $r$, i.e.,
the {\em modified Kepler's law\/} for the binary:
\begin{equation}
\Omega^2 = \frac{2M}{r^3}(1+2 \delta)
=2 \muc (1+2 \delta),      \label{kepler}
\end{equation}
where we have defined
\begin{equation}
\delta\equiv\frac{3}{2}\,\frac{(2I_{11}-I_{22}-I_{33})}{Mr^2},
{}~~~~~\muc \equiv M/r^3.
\label{deltadef}
\end{equation}
The second condition, $\partial E/\partial \rho_c=0$, leads
to the virial relation for the binary,
\begin{equation}
\frac{3}{n}U+W+2T_s=-\frac{2M^2}{R}g_t,      \label{virial}
\end{equation}
where we have defined
\begin{equation}
g_t\equiv \frac{R}{Mr^3}\,(2I_{11}-I_{22}-I_{33})
=\frac{2}{3}\,\frac{R}{r} \delta.
\end{equation}
{}From equations (\ref{Udef}), (\ref{Wdef}) and (\ref{virial}),
the equilibrium mean radius $R$ can be obtained as
\begin{equation}
R=R_o\left[\left(1-2\frac{T_s}{|W|}\right)f
-\left(\frac{5-n}{3}\right)\,g_t\right]^{-n/(3-n)},  \label{RvsR0}
\end{equation}
where $R_o$ is the radius of a spherical equilibrium polytrope with
the same mass and entropy,
\begin{equation}
R_o=\xi_1 \left(\xi_1^2|\theta'_1|\right)^{-(1-n)/(3-n)}
\left[\frac{(n+1)K}{4\pi}\right]^{n/(3-n)}
\left(\frac{M}{4\pi}\right)^{(1-n)/(3-n)}.   \label{Rodef}
\end{equation}

The last two conditions $\partial E/\partial\lone=\partial E/\partial\ltwo=0$,
together with the virial relation, can be used to derive two equations
determining the axes ratios in equilibrium,
\begin{eqnarray}
q_n{\tilde \mu}_R \left[\frac{Q_{1}^2}{\muc}a_2^2
+2 \left(2+2\delta+\frac{Q_{2}\Omega}{\muc}\right)\,a_1^2
+a_3^2\right] &=& 2 (a_1^2A_1-a_3^2A_3),       \label{axis1}   \\
q_n{\tilde \mu}_R \left[ \frac{Q_{2}^2}{\muc}a_1^2
+\left(1+4\delta-\frac{2Q_{1}\Omega}{\muc}\right)\,a_2^2
+a_3^2\right] &=& 2(a_2^2A_2-a_3^2A_3),      \label{axis2}
\end{eqnarray}
where we have defined
\begin{eqnarray}
  Q_1&\equiv&-\frac{a_1^2}{a_1^2+a_2^2}\,\zeta=+\frac{a_1}{a_2}\Lambda, \\
  Q_2&\equiv&+\frac{a_2^2}{a_1^2+a_2^2}\,\zeta=-\frac{a_2}{a_1}\Lambda,
\end{eqnarray}
and $q_n\equiv \kappa_n(1-n/5),~{\tilde \mu}_R\equiv\muc /(\pi \bar \rho)$,
with $\bar \rho= 3M/(4\pi a_1a_2a_3)$.

For any given $M$, $J$, $\cc$, $K$, and $n$, an equilibrium
model can be constructed from the four algebraic equations
(\ref{kepler}), (\ref{RvsR0}), (\ref{axis1}) and (\ref{axis2}).
For specified values of $f_R$ and $r/a_1$, we solve equations
(\ref{axis1}) and (\ref{axis2}) for the axis ratios $a_2/a_1$ and $a_3/a_1$
by a Newton-Raphson method following an initial guess.
The mean radius of the star is then obtained from equation (\ref{RvsR0}).
The total equilibrium energy of the system can be written
\begin{equation}
E_{eq}=2 E_s+\frac{1}{2}\mu r^2\Omega^2-\frac{M^2}{r}
-\left(\frac{2 n+3}{3}\right)\frac{M}{r^3}(2I_{11}-I_{22}-I_{33}),
\label{Eeqdef}
\end{equation}
where $E_s$ is the intrinsic energy of each star,
\begin{equation}
E_s=-\frac{(3-n)M^2}{(5-n)R}f \left[1-\left(\frac{3-2n}{3-n}\right)\,
\frac{T_s}{|W|}\right].
\end{equation}
The total equilibrium angular momentum $J_{eq}$
is given by equation (\ref{Jdef}), evaluated for the equilibrium solution.

\subsection{Equilibrium Sequences with Constant $\cc$}

Since the circulation $\cc$ is conserved in the absence of viscosity,
it is useful to construct sequences of equilibrium configurations
with fixed $\cc$. A constant-$\cc$ sequence is parametrized by the binary
separation $r$ or the angular momentum $J$.
Note that in general $\fch$ varies along a constant-$\cc$ sequence.
When $\cc$ is specified, the value of $f_R$ needs to be determined
simultaneously with $a_2/a_1$ and $a_3/a_1$ to satisfy equation (\ref{Cdef}).

Of particular interest here is the {\em irrotational} Darwin-Riemann
sequence, for which the circulation $\cc=0$ ($f_R=-2$). This corresponds
to the case where the stars have no spin at large separation.
In Table~1 we have listed some properties of these irrotational
Darwin-Riemann configurations for $n=0$, 0.5, 1, and 1.5. All sequences are
terminated when the surfaces of the two stars are in contact, i.e.,
$r/a_1=2$. Following LRS1, we adopt units based on the radius $R_o$
of a spherical equilibrium polytrope with same mass and entropy
(eq.~[\ref{Rodef}]), defining the dimensionless quantities
\begin{equation}
\bar \Omega=\frac{\Omega}{(\pi {\bar\rho}_o)^{1/2}},~~~~
\bar J=\frac{J}{(M^3R_o)^{1/2}},~~~~
\bar E=\frac{E}{(M^2/R_o)},          \label{bardef}
\end{equation}
where ${\bar\rho}_o=M/(4\pi R_o^3/3)$. Note that $R/R_o>1$ for $n>0$,
indicating that the volume of each star increases when placed in a binary
system.

Equilibrium sequences with constant $\cc\ne 0$ can also be constructed.
The value of $\cc$ can be specified from the spin angular frequency
$\Omega_s$ of each star at large separation.
Indeed, for large $r$, we have $a_1\rightarrow a_2$,
$\Omega^2\rightarrow2M/r^3$ and
\begin{eqnarray}
J &\rightarrow & \mu r^2 \Omega -2 I \Lambda
\equiv \mu r^2 \Omega +2 I \Omega_s,  \label{Jlim} \\
\cc &\rightarrow & 2 I \Lambda \equiv -2 I \Omega_s,    \label{Clim}
\end{eqnarray}
where we have identified $\Omega_s=-\Lambda(r=\infty)$
as the spin angular velocity at large $r$
(for an axisymmetric star, uniform spin and vorticity
are indistinguishable in the ellipsoidal models).
Note that when $\Omega_s$ is positive (i.e., the spin is in the same direction
as the orbital angular momentum), $\cc$ is negative.
The maximum spin that a uniformly rotating neutron star
 can sustain without shedding mass from its equator
is given by (Friedman, Ipser \& Parker 1986; Cook, Shapiro \& Teukolsky 1992)
\begin{equation}
{\hat \Omega_s}\equiv {\Omega_s \over (M/R_o^3)^{1/2}} \lo 0.6.
\label{omegmax}
\end{equation}
For typical neutron stars, this maximum spin rate is not very
sensitive to the adopted equation of state.
Table~2 gives the equilibrium properties of the constant-$\cc$ Darwin-Riemann
sequences corresponding to ${\hat\Omega_s}=0.2$ and $0.4$, for
a polytropic index $n=0.5$ or~1. Irrotational
cases, corresponding to $\hat\Omega_s=0$, were given in Table 1.

In Figure~1, we show the variation of the total equilibrium
energy $E_{eq}(r)$ and angular momentum $J_{eq}(r)$ of the binary system,
as well as the orbital angular frequency $\Omega$,
along several sequences with constant
$\cc=-2 I \Omega_s$, for ${\hat\Omega_s}=0,~0.1,~0.2,~0.4$ and
polytropic index $n=0.5$.
Figure~2 shows the variation of the principal axes along the same sequences.
In these plots, each curve terminates at the contact solution.
We see immediately that {\em there exists a critical separation $r_m$
where $E_{eq}(r)$ and $J_{eq}(r)$ are simultaneously minimum\/}.
That $E_{eq}$ and $J_{eq}$ attain their minimum
simultaneously is a consequence of the property
$dE_{eq}=\Omega\, dJ_{eq}$ along an equilibrium sequence with constant $\cc$
(LRS1, Appendix D).
The minimum occurs as a result of the strong
tidal interaction between the two stars at small separation (see Paper~1 for
a qualitative discussion).

{\em The minima in $E_{eq}(r)$ and $J_{eq}(r)$ along a constant-$\cc$ sequence
indicate the onset of dynamical instability\/}.
Indeed, at $r=r_m$, it becomes possible for a small dynamical perturbation
of the system (which conserves $\cc$) to cause no first-order change in
the equilibrium energy or angular momentum.
Such a perturbation must have eigenfrequency $\omega^2=0$, signaling the
onset of instability (see, e.g., Shapiro \& Teukolsky 1983, Chap.~6;
Tassoul 1978).
More rigorously, it can be shown (LRS1) that the onset of instability,
determined from the condition
\begin{equation}
{\rm det}\biggl({\partial^2E\over\partial \alpha_i\partial \alpha_j}
\biggr)_{eq}=0,~~~~i,j=1,2,\ldots~~~~~~({\rm onset~of~instability}),
\end{equation}
where the $\alpha_i$'s are the parameters specifying the
configuration (in this case, $r,\rho_c,\lambda_1$ and $\lambda_2$),
exactly coincides with the points of minimum $E_{eq}(r)$ and $J_{eq}(r)$.
Binary configurations with $r<r_m$ are thus {\em dynamically unstable\/}.
{}From Table~1, we see that $r_m$ is smaller for larger $n$.
This is because tidal effects become important at smaller separation
for more centrally-concentrated stars.
When $n\ga 1.2$, the minimum disappears and all binary configurations
with $\cc=0$ remain stable up to contact.
For sequences with $\cc\ne 0$, we see from Table~2 that,
for a given $n$, the minima become more shallow and
ultimately disappear as $|\cc|$ increases.
Qualitatively, this is because the intrinsic oblateness of the two spinning
stars causes them to come into contact at a larger separation, where
tidal effects are smaller.

\subsection{Synchronized Equilibrium Sequences}

When viscosity is important, the circulation $\cc$ is
no longer conserved during the evolution of the binary.
In the limit where the synchronization timescale is much smaller than
the orbital decay timescale, the evolution of the system may be
described approximately by a sequence of
{\em uniformly rotating\/} (i.e., synchronized) equilibrium configurations.
This is the case for the vast majority of observed binaries, but may not
be true for neutron star binaries (see Bildsten \& Cutler 1992; Kochanek 1992).
Nevertheless, as a limiting case, we also construct equilibrium sequences
with uniform rotation. These represent the compressible generalizations of
the classical incompressible Darwin configurations discussed in
Chandrasekhar (1969). The corotating configurations can be
constructed as a special case of the Darwin-Riemann solutions with
$f_R=0$ and $\Lambda=0$. The total kinetic energy
(eq.~[\ref{T2def}]) in this case simply reduces to
\begin{equation}
T={1\over 2}(\mu r^2+2I)\Omega^2
={J^2 \over 2(\mu r^2+2I)}.
\end{equation}
Extensive tables and plots illustrating the properties of compressible Darwin
sequences are given in LRS1 and will not be repeated here.
Note that the circulation $\cc$ varies along those sequences.

As in Figure~1, it is found that {\em $E_{eq}(r)$ and $J_{eq}(r)$
can attain a minimum at a critical separation $r_m$ before
contact is reached\/}. This occurs for sufficiently incompressible
configurations, with $n\la 2$.
The minimum is more pronounced in this case because of large positive
contributions to the $E_{eq}$ and $J_{eq}$ from the synchronized spins.

Here {\em the minimum marks the onset of secular instability along the
sequence\/}. In the presence of viscosity, configurations with $r<r_m$ will be
driven away from synchronization (see Counselman 1973, and Hut 1980, for
simple models of secularly unstable binaries).
The instability at $r=r_m$ cannot be dynamical
because neighboring configurations along the sequence are still
in uniform rotation and therefore can only be reached on a viscous timescale
(recall that dynamical perturbations conserve $\cc$, which
varies along the corotating sequence).
True dynamical instability occurs a little further along the sequence (at
slightly smaller $r$),
when neighboring configurations {\em with the same value of} $~\cc$ can be
reached with no change in equilibrium energy to first order (see LRS1,
in particular Fig.~14). In this paper (as in Paper~1), for simplicity,
we do not distinguish between the secular and dynamical stability limits
and we treat the instability at $r=r_m$ as if it were dynamical.
This is justified because the binary separation changes very little
between the secular and dynamical stability limits, and departures
from synchronization remain always small.
See LRS2, however, for a detailed discussion of what really happens
between the secular and dynamical stability limits.

\subsection{Models of Black Hole -- Neutron Star Binaries}

Binary systems consisting of a point-like object orbiting a finite-size star,
such as a black hole -- neutron star (BH-NS) binary,
can be modeled as Roche-Riemann
configurations. We have studied such configurations in detail in LRS1.
Compared to the equal-mass Darwin-Riemann configurations
discussed in \S 2.1--2.3, a prominent new feature in the Roche-Riemann systems
is the existence of a {\em Roche limit\/} prior to contact for circular
equilibrium. The Roche limit corresponds to the point at which
the binary separation $r$ has a minimum value below which
no equilibrium solution exists. At the Roche limit, the slope of the
$E_{eq}(r)$ and $J_{eq}(r)$ curves becomes infinite.
Beyond the Roche limit, there exists a second branch
of equilibrium solutions, with larger surface deformation, that extends
all the way to contact.
However, these solutions beyond the Roche limit are unphysical since they
have higher energies than those along the main equilibrium branch for the
same value of $r$ (cf.\ Figs.~10, 11, and~13 of LRS1). When the orbit
of a binary decays to the Roche limit, tidal disruption or mass transfer
is unavoidable.

For Roche-Riemann equilibrium sequences with constant $\cc$,
a dynamical instability is always encountered prior to the Roche
limit. Indeed, the dynamical stability limit corresponds to the minimum
in the $E_{eq}(r)$ curve, which must always precede the point of infinite
slope. Thus {\em binaries at the Roche limit are always dynamically
unstable\/}. In contrast, along a sequence of synchronized configurations
(a compressible Roche sequence), the minimum of
$E_{eq}(r)$ only corresponds to a secular instability, as discussed
in \S 2.3. Dynamical instability sets in later (at smaller $r$)
and the Roche limit may or may not be dynamically unstable,
depending on the mass ratio and the compressibility of the star.
For typical BH-NS binaries with moderate mass ratios
($M_{BH}/M_{NS} \simeq 10$) and highly incompressible neutron star matter
($n \simeq 0.5$), we find that the dynamical
instability sets in {\em prior\/} to the Roche limit (see Table~10 of LRS1).
Consequently, these BH-NS binaries are always dynamically unstable at
the Roche limit. Only when the neutron star is very compressible or when the
black hole
is much more massive, can the Roche limit set in prior to the dynamical
stability limit. In those cases, a dynamically stable binary at the Roche limit
can exist.

In the rest of this paper, we focus on NS-NS binaries, rather than
BH-NS binaries, for the following reason.
The onset of instability (secular or dynamical) along a Roche or
Roche-Riemann sequence occurs at $r_m \sim 2 (1+q)^{1/3} R$,
where $q=M_{BH}/M_{NS}$. But the last stable circular orbit
around a Schwarzschild black hole is at $r_{GR} \sim 6 M_{BH}$
for $M_{BH}\gg M_{NS}$. Thus for a typical BH-NS system with $q \go 10$ and
$R/M_{NS} \simeq 5$, we have $r_{GR}>r_m$
and general relativistic corrections to the orbital motion
are expected to dominate over the Newtonian hydrodynamic effects
discussed here. The situation is different for NS-NS binaries
with $q\simeq1$, where $r_{GR}\la r_m$ typically.
General relativistic corrections to our Newtonian treatment for
NS-NS binaries are discussed in \S 5.

\subsection{Equilibrium Models for Two Unequal Masses}

General Darwin-Riemann models for two nonidentical stars
can be similarly constructed with our energy variational method (see LRS2).
In particular, models for two finite-size stars with
different masses, radii, polytropic indices, adiabatic constants and spins
can be constructed.
Such models can be used to describe binaries containing two
nonidentical neutron stars.
As in Roche-Riemann binaries (\S 2.4), when the masses of the two
stars are sufficiently different, a Roche limit can exist prior to contact,
providing a possibility of mass transfer.
Here also, the Roche limit configuration is always secularly unstable,
but can be dynamically stable or unstable depending on the mass ratio and
compressibility. We return to these questions in \S 6,
where we model neutron star binaries as general Darwin-Riemann
configurations in which the effective polytropic indices and radii of the two
components
are determined from a realistic equation of state and are functions of mass.

\section{ORBITAL EVOLUTION MODEL}

We now wish to study the orbital evolution of the binary models
constructed in \S 2 in the presence of gravitational wave emission.
As in \S 2, we consider binaries in circular orbits. This is
probably justified for most systems at large separation
since gravitational radiation itself tends to
circularize an eccentric orbit. Near contact, however, if relativistic effects
are sufficiently strong, the eccentricity can actually grow again
as the inspiral accelerates (Lincoln \& Will 1988).
This would be the case for two point masses approaching the last stable
circular orbit allowed by general relativity.
For neutron stars with finite
radii, however, we expect hydrodynamic effects to  become important before this
relativity-induced eccentricity can grow significantly.
We will return to this question in \S5.
Here, for simplicity, we assume that the orbit  remains always
circular, at least in some average sense. We determine
the evolution of the average binary separation $r$ as the system loses
energy and angular momentum to gravitational radiation.
Such an approach is clearly valid when the orbital decay time
$t_r=|r/{\dot r}|$ is much larger than orbital period
$P=2\pi/\Omega$. But we  adopt this
approximation here even for the final phase of the orbital decay, when
we find that the two timescales can become comparable.
By doing so, we can study the transition from the secular orbital
decay at large $r$ to the dynamical coalescence at small $r$.
We also assume that the internal structures of the two stars, and
in particular, their shapes, assume the form
for circular equilibrium configurations, i.e., we treat only
$r$ as a dynamical variable, assigning all internal degrees of freedom
($\rho_c$, $\lone$, $\ltwo$) to their equilibrium values.
This approximation is valid as long as the internal dynamical time
(the response time) $t_{dyn}\sim (R^3/M)^{1/2}$
of the stars remains much smaller than the orbital decay time $t_r$,
a condition which is is well satisfied at large separation, and
still marginally satisfied for two neutron stars near contact.

\subsection{Gravitational Radiation}

We calculate the emission of gravitational waves
in the weak-field, slow-motion limit
(see, e.g., Misner, Thorne \& Wheeler 1970).
In this approximation, the rate of energy loss is given by the
usual quadrupole formula, (we set $G=c=1$)
\begin{equation}
\left(\frac{d \ce}{dt}\right)_{GW}
=-\frac{1}{5} \sum_{i,j}
\left\langle \left(\frac{d^3 \Ibar_{ij}}{dt^3}\right)^2
\right\rangle,     \label{quadfor}
\end{equation}
where $\Ibar_{ij}$ is the reduced quadrupole tensor of the system
\begin{equation}
\Ibar_{ij} = \int \rho \left(x_ix_j-\frac{1}{3} {\bf x}^2\delta_{ij}
\right)\,d^3x.     \label{Ibardef}
\end{equation}
For a binary system containing two stars of mass $M$ and $M'$ orbiting
in the $xy$-plane, the only time-dependent components of the quadrupole tensor
are
\begin{eqnarray}
\Ibar_{xx} &=& [\mu r^2 +(I_{11}+I_{11}'-I_{22}-I_{22}')]
\,{1\over 2}\cos\Phi +{\rm constant},  \label{Ixxdef} \\
\Ibar_{xy}=\Ibar_{yx} &=& [\mu r^2 +(I_{11}+I_{11}'-I_{22}-I_{22}')]
\,{1\over 2}\sin\Phi +{\rm constant},    \label{Ixydef}  \\
\Ibar_{yy} &=& -[\mu r^2 +(I_{11}+I_{11}'-I_{22}-I_{22}')]
\,{1\over 2}\sin\Phi +{\rm constant},     \label{Iyydef}
\end{eqnarray}
where $\mu=MM'/(M+M')$ is the reduced mass,
 $I_{ij}$ and $I_{ij}'$ are the quadrupole moments of each star
(cf.\ eq.~[\ref{Iijdef}]), and we have defined an angle
\begin{equation}
\Phi\equiv2 \int^t \Omega\, dt+{\rm constant}.
\end{equation}
Expressions~(\ref{quadfor}) and~(\ref{Ibardef}) then give
\begin{equation}
\left(\frac{d \ce}{dt}\right)_{GW}
=-\frac{32}{5}\,\Omega^6\, (\mu r^2)^2 \,
\biggl [1+{1\over \mu r^2}
(I_{11}+I_{11}'-I_{22}-I_{22}')\biggr]^2.   \label{EdotGW}
\end{equation}
For a circular orbit, the rate of angular momentum loss is given by
(LRS1, Appendix~D)
\begin{equation}
\left(\frac{dJ}{dt}\right)_{GW}=\frac{1}{\Omega}
\left(\frac{d \ce}{dt}\right)_{GW},   \label{JdotGW}
\end{equation}
while the fluid circulation $\cc$ is strictly conserved (Miller 1974).
In expression (\ref{EdotGW}), the second term in the bracket
represents the correction to the point-mass result due to tidal effects.
For large $r$, this is a small correction, of the order of
$\kappa_n (R/r)^5$ (Clark 1977), but it can become as much as
$\sim 40 \%$ near contact.

In the quadrupole approximation, the wave
amplitude $h_{ij}^{TT}$ in the transverse-traceless (TT) gauge is given by
\begin{equation}
h_{ij}^{TT}=\frac{2}{D}\,{\ddot\Ibar}_{ij}^{{}\,TT}(t-D),
\end{equation}
where $\Ibar_{ij}^{TT}$ is the transverse projection
of the reduced quadrupole moment,
$D$ is the distance between source and observer, $t-D$ is the retarded
time, and a dot indicates a time derivative.
For wave propagation in the direction $(\theta,\phi)$ in
spherical coordinates with orthonormal basis vectors
${\bf e}_{\hat r}$, ${\bf e}_{\hat\theta}$, and ${\bf e}_{\hat\phi}$,
the two basis polarization tensors are
\begin{eqnarray}
{\bf e}_{+} &=& {\bf e}_{\hat\theta} \otimes {\bf e}_{\hat\theta}
  -{\bf e}_{\hat\phi} \otimes {\bf e}_{\hat\phi},\\
{\bf e}_{\times} &=& {\bf e}_{\hat\theta} \otimes {\bf e}_{\hat\phi}
  +{\bf e}_{\hat\phi} \otimes {\bf e}_{\hat\theta}.
\end{eqnarray}
In this basis, $h_{ij}^{TT}$ can be written
\begin{equation}
{\bf h}^{TT}\equiv h_{+}{\bf e}_{+}+h_{\times}{\bf e}_{\times}
 =\frac{1}{D}({\ddot\Ibar}_{\hat\theta\hat\theta}
  -{\ddot\Ibar}_{\hat\phi\hat\phi})\,{\bf e}_{+}+
   \frac{2}{D}{\ddot\Ibar}_{\hat\theta\hat\phi}\,{\bf e}_{\times},
\label{hTTdef}
\end{equation}
where the spherical components of the reduced quadrupole tensor are
given by (we set $\phi=0$ without loss of generality)
\begin{equation}
{\ddot\Ibar}_{{\hat\theta}{\hat\theta}}={\ddot\Ibar}_{xx}\cos^2\theta,~~~~~~
{\ddot\Ibar}_{{\hat\phi}{\hat\phi}}={\ddot\Ibar}_{yy},~~~~~~
{\ddot\Ibar}_{{\hat\theta}{\hat\phi}}={\ddot\Ibar}_{xy}\cos\theta.
\label{Iddot}
\end{equation}
Using equations (\ref{Ixxdef})---(\ref{Iyydef}),
(\ref{hTTdef}) and~(\ref{Iddot}),
we find the following expressions for the waveforms
\begin{eqnarray}
h_{+} &=& -\frac{2}{D}\Omega^2 [\mu r^2 +(I_{11}+I_{11}'-I_{22}-I_{22}')]
(1+\cos^2\theta)\cos\Phi, \label{hplus} \\
h_{\times} &=& -\frac{4}{D}\Omega^2 [\mu r^2 +(I_{11}+I_{11}'-I_{22}-I_{22}')]
\cos\theta \sin\Phi. \label{hcross}
\end{eqnarray}

Note that the above derivation neglects the contribution from the
orbital decay itself to the gravitational radiation.
The correction to the wave amplitude due to the orbital decay is
of order $|\dot r/(r\Omega)|$, and the correction to
the energy loss rate is of order ${\dot r}^2/(r\Omega)^2\sim (M/r)^5$. This is
smaller by a factor $\sim(M/R_o)^5$ than the correction due to
tidal effects that we have included in expression~(\ref{EdotGW}).

\subsection{Orbital Evolution at Large Separation}

For sufficiently large orbital separation, we expect the orbital
decay to proceed  quasi-statically (i.e., along an equilibrium
sequence) with the rate of change ${\dot r}$ of the orbital
separation given by
\begin{equation}
\dot r=\left(\frac{d \ce}{dt}\right)_{GW}
\left(\frac{d E_{eq}}{dr}\right)^{-1}.    \label{rdot}
\end{equation}
This expression must obviously break down when $E_{eq}(r)$ is
near a minimum, since it would otherwise predict
that ${\dot r}\rightarrow\infty$ there.
In this section we explore the effects of tidal interactions in the limit of
large $r$, when expression (\ref{rdot}) applies. In particular,
we calculate analytically the deviations from the point-mass behavior due
to the finite-size effects.
In \S 3.3, we will develop and implement a numerical formalism
(ODE's) to calculate the orbital
decay at smaller $r$, when equation~(\ref{rdot}) no longer applies.

\subsubsection{Point-Mass Results}

For binaries containing two point masses $M$ and $M'$
we have $E_{eq}=-MM'/(2r)$
and $I_{ij}=I_{ij}'=0$ in equation (\ref{EdotGW}),
so that equation (\ref{rdot}) yields the familiar result
\begin{equation}
{\dot r}=-\frac{64}{5}{\mu M_t^2\over r^3},  \label{rodot}
\end{equation}
where $M_t=M+M'$. The orbital evolution is obtained by integration,
\begin{equation}
[r(t)]^4={256\over 5}\mu M_t^2 (T-t), \label{rt}
\end{equation}
where $T$ is the time at the end of the coalescence ($r=0$).
The frequency and phase of the gravitational waves are then easily obtained as
\begin{eqnarray}
f_{GW}(t) &=& {\Omega\over\pi}={1\over\pi}\biggl[{5\over 256}
{1\over\mu M_t^{2/3}(T-t)}\biggr]^{3/8},\\
\Phi(t) &=& \Phi_o -2 \biggl({T-t\over 5\mu^{3/5}M_t^{2/5}}\biggr)^{5/8},
\end{eqnarray}
where $\Phi_o$ is a constant and we have used $\Omega^2=M_t/r^3$,
the Keplerian value.

Of great importance for the detection of gravitational
waves by laser interferometers is the number of orbits $N_{orb}$ (or the
number of cycles of gravitational waves $N_{GW}=2N_{orb}$)
in a given interval of wave frequency or binary separation
(Cutler et al.~1993). This is obtained by integrating
\begin{equation}
dN_{orb}=\frac{\Omega}{2\pi}dt = {\Omega\over 2\pi}
\biggl({dE_{eq}\over dr}\biggr)
\biggl({d\ce\over dt}\biggr)_{GW}^{-1} dr.   \label{dNorbdef}
\end{equation}
For point masses, this reduces to
\begin{equation}
dN_{orb}^{(0)}=-{5\over 128\pi\mu M_t^{3/2}}r^{3/2} \,dr
 ={5\over 192\pi\mu M_t^{2/3}}{1\over (\pi f_{GW})^{5/3}}
  d(\ln f_{GW}).   \label{dNorb0}
\end{equation}
Integrating equation~(\ref{dNorb0}) we find
the number of orbits between $r_i$ and $r_f<r_i$,
\begin{equation}
N_{orb}^{(0)} = \frac{1}{64\pi} \frac{(r_i^{5/2}-r_f^{5/2})}
 {MM'M_t^{1/2}}.    \label{Norb0}
\end{equation}
Any deviation from this result leading to a change $\delta N_{orb}\go0.06$
over the detection interval (corresponding to a phase change
$\delta\Phi=2\pi\delta N_{GW}=4\pi\delta N_{orb}\go\pi/4$)
must be incorporated into the theoretical waveform templates used
for signal extraction.

We now discuss the deviations from the result~(\ref{Norb0}) caused by
finite-size effects. We consider the three types of binary configurations
introduced in \S 2: irrotational ($\cc=0$) configurations (\S 3.2.2),
configurations with $\cc=\,$constant$\ne0$ (\S 3.2.3), and synchronized
configurations (\S 3.2.3).
For simplicity, we assume that $M'$ is a
point mass (i.e., we consider only Roche and Roche-Riemann binaries),
but the generalization to two stars of finite size is straightforward
in this case: one would simply add another contribution obtained by
interchanging $M$ and $M'$ in the results given below.

\subsubsection{Irrotational Configurations}

For nonspinning neutron stars ($\cc=0$),
the correction to $N_{orb}^{(0)}$ at large separation is entirely
due to the tidal interaction. Since the tidally induced ellipticity
of the star is $\sim (R_o/r)^3$,
the tidal corrections to $\Omega$, $(d\ce/dt)_{GW}$, and $dE_{eq}/dr$
are all of order $(R_o/r)^5$. Thus from equation~(\ref{dNorbdef}), we expect
a change $\delta (dN_{orb}^{(I)})\propto (R_o/r)^5 dN_{orb}^{(0)}
\propto r^{-7/2}dr$. A detailed calculation
based on series expansions of the binary equilibrium equations at large $r$
(see LRS2) gives for the total equilibrium energy
\begin{equation}
E_{eq}(r)=-\frac{MM'}{2r}
  +\frac{3}{2}\kappa_n q_n M'^2 \frac{R_o^5}{r^6}
  +\ldots \,\, +{\rm constant},
\end{equation}
(recall that $q_n=\kappa_n(1-n/5)$ and $\kappa_n$ is
given by eq.~[\ref{kndef}])
and for the orbital angular frequency
\begin{equation}
\Omega^2={M+M'\over r^3}\left(1+
  \frac{9}{2}\kappa_nq_n\frac{M'}{M}\frac{R_o^5}{r^5} +\ldots\right).
\end{equation}
The correction to $(d\ce/dt)_{GW}$ due to the quadrupole moment of $M$
(cf.\ eq.~[\ref{EdotGW}]) is
\begin{equation}
\frac{(I_{11}-I_{22})}{\mu r^2}=\frac{3}{2}\kappa_n q_n
  \frac{M_t}{M}\frac{R_o^5}{r^5} +\ldots
\end{equation}
Equation~(\ref{dNorbdef}) then gives the change in $N_{orb}$ for
irrotational configurations\footnote{The numerical coefficients in
our result do not agree with those
given by Kochanek 1992 (cf.\ his eq.~[5.4]).}
\begin{equation}
\delta N_{orb}^{(I)} = -\frac{3}{64\pi} \kappa_n q_n
  \left(\frac{39}{4} + \frac{M_t}{M'}\right)
   R_o^5\,\frac{(r_f^{-5/2}-r_i^{-5/2})}{M^2M_t^{1/2}}.  \label{dNorbI}
\end{equation}
We see clearly that this deviation from the point-mass result
does not accumulate at large $r$. The term proportional to $M^{-2}M_t^{-1/2}$
in expression~(\ref{dNorbI}) results from the change in $\Omega$
and $E_{eq}$, while the
term containing the extra factor of $M_t/M'$ comes from the increase in
$(d\ce/dt)_{GW}$ caused by the quadrupole moment of the star.

\subsubsection{Configurations with constant $\cc\ne0$}

For neutron stars with nonzero spin, the dominant effect at large $r$ is
the change in $\Omega$ caused by the
spin-induced quadrupole moment (Bildsten \& Cutler 1992).
Indeed, for sufficiently large $r$, tidal effects
can be ignored, and the star can be modeled as an
axisymmetric compressible Maclaurin spheroid, with $I_{11}=I_{22}>I_{33}$.
The quadrupole interaction energy is $-M'(I_{11}-I_{33})/(2r^3)$
and we find (LRS2) that the total equilibrium energy can be written simply as
\begin{equation}
E_{eq}(r)=-{MM'\over 2r}+{M'\over 4r^3}(I_{11}-I_{33})\,\,+{\rm constant},
\end{equation}
while the orbital angular frequency is given by
\begin{equation}
\Omega^2={M+M'\over r^3}\biggl[1+{3\over 2Mr^2}(I_{11}-I_{33})\biggr].
\end{equation}
The correction factor for $(d\ce/dt)_{GW}$ is then simply
$[1+3(I_{11}-I_{33})/(2Mr^2)]^3$. Therefore from
equation~(\ref{dNorbdef}) we have
\begin{equation}
dN_{orb}= dN_{orb}^{(0)}\,\biggl[1-{3\over 2Mr^2}(I_{11}-I_{33})\biggr]
\biggl[1+{3\over 2Mr^2}(I_{11}-I_{33})\biggr]^{-5/2}.
\end{equation}
At large $r$ this gives for the change in $N_{orb}$ due to spin:
\begin{equation}
\delta N_{orb}^{(S)} = -\frac{105}{256\pi}\,\frac{(I_{11}-I_{33})}{M^2M'}\,
\frac{(r_i^{1/2}-r_f^{1/2})}{M_t^{1/2}}.   \label{dNorbS}
\end{equation}
This result agrees with equation~(10) of Bildsten \& Cutler (1992).
Expressions for the quadrupole moments $I_{11}$ and $I_{33}$ of
compressible Maclaurin spheroids can be found in LRS1.

If the spin-induced eccentricity $e^2=1-a_3^2/a_1^2 \ll 1$, then
the $\Omega_s(e)$ relation for a rotating spheroid (e.g., LRS1, eq.~[3.21])
can be expanded to give $e^2 \simeq (5q_n/2){\hat\Omega_s}^2$, with
 ${\hat \Omega_s}^2=\Omega_s^2/(M/R_o^3)$. Thus we have:
\begin{equation}
{I_{11}-I_{33}\over MR_o^2}={\kappa_n\over 5}{a_1^2-a_3^2\over R_o^2}
\simeq {\kappa_n\over 5}e^2
\simeq \frac{\kappa_n q_n}{2}{\hat\Omega_s}^2.
\end{equation}
Therefore equation~(\ref{dNorbS}) becomes
\begin{equation}
\delta N_{orb}^{(S)} = -\frac{105}{512\pi} \kappa_n q_n
  \frac{R_o^2}{MM'}\,{\hat\Omega}_s^2
 \frac{(r_i^{1/2}-r_f^{1/2})}{M_t^{1/2}},~~~~~~~~~~~({\hat\Omega}_s^2 \ll 1).
  \label{dNorbSs}
\end{equation}
Since $\delta N_{orb}^{(S)} \propto r^{1/2}$, we see
that this deviation from the point mass result, in contrast to
$\delta N_{orb}^{(I)}$, does accumulate at large $r$.


\subsubsection{Synchronized Configurations}

For the corotating case, the dominant effect is simply the added kinetic energy
and
angular momentum of the synchronized spin (Bildsten \& Cutler 1992; Kochanek
1992).
Expanding equation~(\ref{Eeqdef}) at large $r$, we obtain
\begin{equation}
E_{eq}(r)= -{MM'\over 2r}
+{1\over 5}\kappa_n MR_o^2\,\biggl({M+M'\over r^3}\biggr)
+\ldots\,\,  +{\rm constant}.
\end{equation}
The corrections to $(d\ce/dt)_{GW}$ and $\Omega$ are of higher order
in $R_o/r$, and we find from equation~(\ref{dNorbdef}) that for
synchronized spin,
\begin{equation}
\delta N_{orb}^{(SS)} = -\frac{3}{32\pi} \kappa_n
  \frac{M_t^{1/2}R_o^2(r_i^{1/2}-r_f^{1/2})}{MM'^2}   \label{dNorbSS}
\end{equation}
This result agrees with equation~(3) of Bildsten \& Cutler (1992)
in the incompressible limit (where $\kappa_n=1$).
Here also, since $\delta N_{orb}^{(SS)} \propto r^{1/2}$,
we see that the phase error  accumulates at large $r$.

\subsection{Approach to Dynamical Instability}

Whenever $E_{eq}(r)$ has a minimum at some $r=r_m$,
the orbital decay cannot remain quasi-static as $r_m$ is
approached, since equation~(\ref{rdot}) would predict
that ${\dot r}\rightarrow\infty$ as $r\rightarrow r_m$.
This should not be too surprising since, as discussed
in \S 2, the binary orbit becomes dynamically unstable for
$r<r_m$. As the stability limit is approached, the radial
infall velocity $\dot r$ can become much larger than
predicted for two point masses (eq.~[\ref{rodot}]). For
$r<r_m$, the coalescence would proceed on a dynamical timescale
even in the absence of energy and angular momentum losses.

Let us first estimate the separation $r_c>r_m$ where equation (\ref{rdot})
starts to break down. We can expand $E_{eq}(r)$ around the minimum
at $r=r_m$ as
\begin{equation}
E_{eq}(r)=E_{eq}(r_m)+\frac{\eps}{r_m} (r-r_m)^2+\ldots,~~~~
{}~~~~~~\eps \sim M^2/r_m.   \label{Eeqexp}
\end{equation}
Equation (\ref{rdot}) becomes invalid when
the rate of increase of infall kinetic
energy, becomes comparable to the rate of change of
the equilibrium energy, i.e., $r_c$ is determined by
\begin{equation}
\mu {\dot r} {\ddot r} \sim \left(\frac{dE_{eq}}{dr}\right){\dot r},
{}~~~~{\rm for}~~~~r\sim r_c.
\end{equation}
Using equations (\ref{rdot}) and (\ref{Eeqexp}), this reduces to
\begin{equation}
\mu {\dot \ce}_{GW}{2 \eps \over r_m^2} \sim
\biggl ({dE_{eq}\over dr}\biggr)^4 \sim \biggl ({2\eps\over r_m}\biggr)^4
\delta_c^4, \label{muedot}
\end{equation}
where we have defined
\begin{equation}
\delta_c \equiv (r_c-r_m)/r_m < 1.
\end{equation}
Using equations (\ref{EdotGW}) and (\ref{muedot}), we then obtain
\begin{equation}
\delta_c \equiv \frac{r_c-r_m}{r_m} \sim \left(\frac{r_m}{M}\right)^{-5/4}.
\end{equation}
Even for neutron stars, we have $r_m\simeq3R$, so that $r_m/M \sim 5$
and $r_c$ is not very far from $r_m$, typically 10\% further out.

To properly calculate the orbital evolution for $r<r_c$, when the kinetic
energy of the radial infall becomes important, we write the
total energy of the system (not necessarily in equilibrium), as
\begin{equation}
\ce =\frac{1}{2}\mu{\dot r}^2+E(r,\alpha_i';M,J,\cc),   \label{cedef}
\end{equation}
where the second term is given by equation (\ref{Edef}), and
($\alpha_i'$) denotes the three variables ($\rho_c,\lambda_1,\lambda_2$).
In writing down equation (\ref{cedef}), we have implicitly ignored other
possible contributions to the kinetic energy such as terms
like $M{\dot a_1}^2$ which are related to the change of the structure
of the stars as the binary separation decreases. This is justified
since the adjustment of the stellar shape takes place on the
internal dynamical timescale $t_{dyn}\sim (R_o^3/M)^{1/2}$ of the star, while
the orbital evolution of the binary usually takes place over
a longer time $t_r$. We also assume that the orbit remains quasi-circular, so
that equations (\ref{EdotGW}) and (\ref{JdotGW}) hold. In doing so we
neglect terms of order $e^2$, where $e$ is the eccentricity. Since
$e^2\sim (\dot r/\Omega r)^2$, this is valid so long as $t_r\gg P$. We
find in \S 4 that the combined condition $t_{dyn}\ll P\ll t_r$ is well
satisfied at large $r$, and remains marginally true even when $r$ is
close to $r_m$.

Taking the time derivative of equation (\ref{cedef}),
and recalling that gravitational radiation conserves $\cc$, we get
\begin{equation}
\dot \ce =\mu{\dot r}{\ddot r}+\Omega{\dot J}
+{\partial E\over \partial r}{\dot r}
+\sum_i{\partial E\over \partial\alpha_i'}{\dot\alpha_i'},  \label{de1}
\end{equation}
where we have used $\Omega=\partial E/\partial J$. We now impose
the assumption that the three parameters
$\rho_c,\lambda_1,\lambda_2$ specifying the internal structure of the stars
take their equilibrium values. Accordingly, we have
\begin{equation}
{\partial E\over \partial \alpha_i'}=0,~~~~{\rm for}~~~~
\alpha_i'=(\alpha_i')_{eq}.   \label{pepa}
\end{equation}
Now since $\dot \ce=(\dot\ce)_{GW}$ and $\dot J=(\dot J)_{GW}$,
using equations (\ref{JdotGW}) and (\ref{pepa}),
the evolution equation (\ref{de1}) reduces to
\begin{equation}
\mu \ddot r+\frac{\partial E}{\partial r}=0.
\end{equation}
We now substitute expression (\ref{Edef}) for $E$ and find on differentiating
\begin{equation}
\ddot r -\Omega^2r+\Omega_{eq}^2r=0,~~~~{\rm with}
{}~~~~\Omega_{eq}^2= \frac{2M}{r^3}+\frac{6}{r^5}(2I_{11}-I_{22}-I_{33}).
\label{rddot}
\end{equation}
In the last expression, the $I_{ij}$ assume their equilibrium values as
a function of $r$. We now express $\Omega$ in terms of $J$ using
equations~(\ref{Jdef})-(\ref{Cdef}). After some algebra we get
\begin{equation}
\Omega=\frac{J}{I_t(r)}+F(r),  \label{Omdef}
\end{equation}
where
\begin{equation}
F(r)= \frac{4\cc}{I_t}\frac{a_1a_2}{a_1^2+a_2^2},~~~~~
I_t=\mu r^2+\frac{2}{5}M\kappa_n\frac{(a_1^2-a_2^2)^2}{a_1^2+a_2^2}.
\label{Itdef}
\end{equation}

Equations (\ref{rddot})---(\ref{Itdef}), together with
equations~(\ref{EdotGW}) and~(\ref{JdotGW}) for ${\dot J}
=(dJ/dt)_{GW}$ can be integrated numerically given initial conditions at
any separation $r_i$ such that $(r_i-r_m)/r_m\gg \delta_c$.
We cast these equations into a set of first order ODE's for
${\dot r},~{\dot v_r}={\ddot r}$ and $\dot J$. We
calculate initial values for $\dot r$ and $\ddot r$ at $r=r_i$
from equation~(\ref{rdot}) and then substitute $\ddot r$ into
equation~(\ref{rddot}) to obtain the initial value of $J$.

Alternatively, we can eliminate $J$ completely to obtain a single,
third-order evolution equation for $r$. Taking the time derivative of
equation~(\ref{rddot}) and substituting expression~(\ref{JdotGW})
for ${\dot J}$, we get:
\begin{equation}
\frac{d^3r}{dt^3}-\frac{\dot r\ddot r}{r}\left(1-\frac{2r}{I_t}
\frac{dI_t}{dr}\right)+2r\dot r\left(\frac{dF}{dr}+\frac{F}{I_t}
\frac{dI_t}{dr}\right)(\Omega_{eq}-\Omega)
+\frac{2r}{I_t}\frac{dE_{eq}}{dr}\dot r=\frac{2r}{I_t}
\left(\frac{d\ce}{dt}\right)_{GW},    \label{rdddot}
\end{equation}
where we have used $\Omega{\dot J}={\dot E}$ and $\Omega_{eq}(dJ_{eq}/dr)=
dE_{eq}/dr$. A similar type of equation was derived by Lattimer \&
Schramm (1976) in their study of tidal disruption by black holes.
For large $r$, the first three terms in equation~(\ref{rdddot})
can be ignored since the acceleration of radial infall is small
(note that $(\Omega_{eq}-\Omega) \propto \ddot r$ from eq.~[\ref{rddot}]),
so that equation~(\ref{rdddot}) reduces simply to
$\dot r(dE_{eq}/dr)=\dot E$, i.e., equation~(\ref{rdot}).

The derivation presented above assumed constant fluid circulation and
is valid in the limit of zero viscosity. In the opposite limit, when
viscosity is always acting on a timescale shorter than the orbital
decay timescale, the binary remains synchronized throughout the
evolution. The energy functional for Darwin (uniformly rotating)
configurations should
then be used in equation~(\ref{cedef}). We find that the orbital
evolution in this case is still given by
equations~(\ref{rddot})--(\ref{Omdef}), but with
\begin{equation}
F=0,~~~~I_t=\mu r^2+2I=\mu r^2+\frac{2}{5}M\kappa_n(a_1^2+a_2^2).
\end{equation}
We reemphasize that the assumption of uniform rotation must break down
for $r<r_m$ since viscosity will then drive the
system toward lower energy, hence, {\em away from\/}, rather than
towards, a synchronized state. Nevertheless, for simplicity, we calculate
the orbital evolution in this limiting case based on the energy of Darwin
configurations, even for $r<r_m$.

\section{APPLICATIONS TO BINARY NEUTRON STARS}

We now apply the orbital evolution model developed in \S3 to  calculate
the coalescence of two neutron stars. In \S4.1, we first discuss how
polytropes can best be used to approximate the internal structure of
a neutron star. In \S 4.2, we examine the effects of the spin and tidal
interaction on the gravitational waveforms at large separation, using the
analytic results of \S 3.2.
We then proceed to solve the orbital evolution equations derived in \S 3.3
and study the approach to dynamical instability at small separation (\S 4.3).

\subsection{Polytropic Models for Neutron Stars}

The main parameter that enters the
evolution equations of \S 3 is the ratio $R_o/M$,
which for neutron stars is determined from the nuclear
equation of state (EOS). For the canonical neutron star mass
$M=1.4 M_{\odot}$, all EOS tabulated in Arnett \& Bowers (1977)
give $R_o/M$ in the range of 4--8. Small values of
$R_o/M$ correspond to a soft EOS, while large values correspond
to a stiff EOS. Very soft EOS, such as that of Friedman \& Pandaripande
(1981), appear have been ruled out by the constraint
on the lower mass limit, $1.55 M_{\odot}$, of the X-ray binary 4U0900-40
(Joss \& Rappaport 1984)
as well as by the constraint derived from pulsar timing after glitches
(Link, Epstein \& Van Riper 1992).
The most recent microscopic EOS constrained by nucleon scattering data
and the binding of light nuclei are those of Wiringa, Fiks \& Fabrocini
(1988, hereafter WFF; see also Baym 1991
for a review). For $M=1.4 M_{\odot}$, the WFF EOS give
$R_o/M \simeq 4.7-5.3$, corresponding to $R_0 \simeq 10.4$--11.2$\,$km.
In addition, the radius is almost independent of the mass for
$M$ in the range of $0.8M_{\odot}$ to $1.5M_{\odot}$. Thus the value of
$R_o/M$ can be somewhat larger for smaller $M$.
In this section we consider $R_o/M=5$ and $R_o/M=8$
as representative values.

A polytrope is only an approximate parametrization for a real EOS.
To find the approximate polytropic index $n$ that mimics
the structure of a real neutron star, we proceed as follows.
For given $M$ and $R_o$, we determine the ratio $I/I_u$, where $I$ is the
moment of inertia tabulated for a real EOS,  and $I_u$ is that of a uniform
sphere with same $M$ and $R_o$. Note that when we include general
relativistic corrections, $I_u$ can be larger than $2MR_o^2/5$.
We calculate $I_u$ using equations (3.8)-(3.11) of Arnett \& Bowers (1977).
The resulting ratio is set equal to $\kappa_n$, from which the
corresponding value of $n$ is obtained. In Table~3, we
list the results for different masses based on the EOS AV14+UVII of WFF.
For the other two EOS given in WFF, the results are very similar.
Typically, for $M\simeq 1.4 M_{\odot}$, we find $n\simeq 0.5$, highly
incompressible.
As $M$ decreases, $n$ increases and the configurations become more
compressible. In the orbital evolution calculations
presented below, we consider
the representative values $n=0.5$ and $n=1$.
We also give some results with $n=1.5$ for comparison.

When translating from dimensionless quantities to physical quantities
such as a gravitational wave frequency in Hz, the values of both
$M$ and $R_o/M$ must be specified. The masses of neutron stars
have been determined for a number of binary radio pulsars
as well as binary X-ray pulsars
(see Thorsett et al.~1993 and references therein).
All the measurements are consistent with a mass $M=(1.35\pm 0.1) M_{\odot}$.
Hence we shall focus on this canonical value of $M=1.4 M_{\odot}$ when we
quote actual wave frequencies.

The initial neutron star spins also enter the calculation in the
zero-viscosity case (cf.\ eq.~[\ref{Clim}]).
In the four neutron star binaries known in our Galaxy, the radio
pulsars have spin periods above $30\,$ms, but much shorter pulsar periods
$\sim 1.5\,$ms have been observed in other systems.
The minimum spin period corresponding to equation (\ref{omegmax}) is
about $0.85\,$ms for $M=1.4\,M_{\odot}$.
Here we consider the representative values
${\hat\Omega}_s=0,~0.1,~0.2$, and 0.4, corresponding to
no spin or spin periods of $4.8\,$ms, $2.4\,$ms, and $1.2\,$ms,
respectively. We assume for simplicity that both stars have the same spin.

\subsection{Orbital Evolution of Binary Neutron Stars at Large Separation}

The expressions derived in \S 3.2 can be evaluated easily for neutron
star binaries.
For definiteness, we consider the typical case where $M=M'=1.4\,M_{\odot}$,
$R_o/M=5$ and $n=0.5$ (the corresponding value of $\kappa_n=0.8148$).
We focus on the constant-$\cc$ evolutions and assume that both stars
have the same spin. The gravitational wave frequency $f_{GW}$ in real
units is given at large $r$ by
\begin{equation}
f_{GW}=5837\,\,M_{1.4}^{-1}\,\left(\frac{R_o}{5M}\right)^{-3/2}
  \,\left(\frac{r}{R_o}\right)^{-3/2}~{\rm Hz},   \label{fGW}
\end{equation}
where $M_{1.4}\equiv M/(1.4M_{\odot})$.
For these parameters, the low-frequency band of interest to LIGO
corresponds approximately to binary separations between
$r_i=70 R_o$ ($f_{GW,i}=10\,$Hz) and $r_f=5 R_o$ ($f_{GW,f}=522\,$Hz).
When $r\lo5R_o$, the analytic results of \S3.2 become inaccurate.
The orbital evolution for $r<5R_o$ is calculated in \S 4.3 using the
method introduced in \S3.3.
The total number of cycles of gravitational radiation
emitted by two {\em point masses\/} between $r_i$ and $r_f$ is
$N_{GW}^{(0)}=16098$ (eq.~[\ref{Norb0}]).

Using equation (\ref{dNorbI}) we can calculate the
accumulated change in $N_{GW}=2N_{orb}$
due to tidal effects in irrotational configurations,
\begin{eqnarray}
\delta N^{(I)}_{GW} &\simeq& -16.5 \,\biggl({R_o\over 5M}\biggr)^{5/2}
\biggl[\biggl(
{R_o\over r}\biggr)^{5/2}-\biggl({R_o\over r_i}\biggr)^{5/2}\biggr]\nonumber\\
&\simeq& -8.74 \times 10^{-6} M_{1.4}^{5/3}\biggl({R_o\over 5M}\biggr)^5
\left(f_{GW}^{5/3}-f_{GW,i}^{5/3}\right),  \label{dni}
\end{eqnarray}
where $f_{GW}$ is the wave frequency in Hz.
This result is illustrated in Figure~3. Note that we have multiplied
equation~(\ref{dNorbI}) by~2 to account for the presence of two identical
stars.
The final change is $\delta N^{(I)}_{GW}(r_f)\simeq 0.30$.
Note that this change accumulates mainly at large $f_{GW}$ (small $r$).
For $f_{GW}<300\,$Hz, corresponding to $r > 7.23R_o$, or the first
16065 cycles, we find $\delta N^{(I)}_{GW}\lo 0.1$, while
between $f_{GW}=300\,$Hz and $f_{GW}=522\,$Hz (the remaining
33 cycles) we find $\delta N^{(I)}_{GW}\simeq 0.2$.
Although these changes are small and can probably be neglected in the
construction of theoretical low-frequency wave templates,
they may nevertheless be detectable by advanced LIGO detectors.

Now turn to the case where the stars are spinning. We assume
that $\hat\Omega_s^2\ll1$ and use expression~(\ref{dNorbSs}) (multiplied by~2
for two stars with the same spin) to get
\begin{eqnarray}
\delta N^{(S)}_{GW} &\simeq& -6.16\, \biggl({R_o\over 5M}\biggr)^{5/2}
\hat\Omega_s^2 \biggl[\biggl(
{r_i\over R_o}\biggr)^{1/2}-\biggl({r\over R_o}\biggr)^{1/2}\biggr]\nonumber\\
&\simeq& -111\, M_{1.4}^{-1/3}\biggl({R_o\over 5M}\biggr)^2
\hat\Omega_s^2 \left(f_{GW,i}^{-1/3}-f_{GW}^{-1/3}\right).  \label{dns}
\end{eqnarray}
The total change is $\delta N^{(S)}_{GW}(r_f)\simeq 37.8\,\hat\Omega_s^2
\simeq 8.85/P_{ms}^2$, where $P_{ms}$ is the spin period in milliseconds.
Clearly, this spin-induced change accumulates mainly at large $r$ (see Fig.~3),
and this effect can be very important for rapidly spinning neutron stars:
to get $\delta N^{(S)}_{GW}(r_f)\go 0.1$, we need a spin period $P_{ms}\lo 9$.
In agreement with Bildsten \& Cutler (1992), we conclude that
finite-size effects
for these rapidly spinning neutron stars are potentially very important
in modeling the gravitational radiation waveforms, even at low frequency (large
$r$).

For synchronized binaries, the change $\delta N^{(SS)}_{GW}$ calculated
from expression~(\ref{dNorbSS}) is much larger (by about two orders of
magnitude) than either $\delta N^{(I)}_{GW}$ or $\delta N^{(S)}_{GW}$.
However, synchronized or nearly-synchronized configurations are particularly
unlikely for binary neutron stars at large $r$. This is because the
ratio of synchronization to orbital decay timescales {\em increases\/} with $r$
(Kochanek 1992). If viscosity plays any role at all during the coalescence,
it is much more likely to be during the final phase.

\subsection{Orbital Evolution of Binary Neutron Stars Near Contact}

We calculate the orbital evolution at small $r$ by integrating
the evolution equations (\ref{rddot})--(\ref{Itdef}) numerically using
a fourth-order Runge-Kutta method with adaptive stepsize to ensure
accuracy. The integration is terminated at the separation $r=r_f$ where a
contact configuration is reached along the equilibrium sequence.
For two identical, slowly spinning neutron stars with mass $M\go0.7$
and effective polytropic index $n\lo1$ (cf.\ Table~3), we always have
$r_f<r_m$, where $r_m$ is the separation where $E_{eq}(r)$ is minimum
(the stability limit, cf.\ \S2).
Since most of our assumptions in both \S2 and \S3 are only
marginally valid when $r<r_m$, the results for $r_f<r<r_m$ should be
considered very approximate.

Figure 4 shows the evolution of a system with $n=0.5$ and $R_o/M=5$.
Constant-$\cc$ evolutions with different values of the initial spin
are shown. The point-mass results (eqs.~[\ref{rodot}], [\ref{rt}],
and~[\ref{dNorb0}])
are also shown for comparison, as well as the results for a corotating system.
We see that, as the stability limit at $r=r_m$ is approached,
the orbital decay rapidly accelerates and departs from the point-mass result,
with $\dot r$ reaching typically about 10\% of the
orbital velocity near contact. As a result, it takes only one more orbit
to complete the final decay from $r_m$ to $r_f$.
We see also in Figure~4 that the results for $\cc \ne 0$
are not very different from those for $\cc=0$. This comes about
from two opposite effects:
the initial spin tends to reduce the instability at small $r$ (see \S 2,
Table~2), while it can accelerate the orbital decay at larger
$r$ due to the spin-induced quadrupole interaction (see \S 3.2.3).
By contrast, we see that the corotating evolution is very
different from the constant-$\cc$ evolution.

Figure 5 compares the solutions for different $n$, all with $R_o/M=5$.
Clearly, for larger $n$ (higher compressibility),
the effects of the instability tend to be smaller. This is because
a more compressible configuration is more centrally concentrated, and
tidal effects only become important
at smaller $r$. As a result both $r_m$ and $r_m-r_f$
(the ``acceleration distance'') are smaller (cf.\ Table~1).

During the evolution, the total angular momentum and  energy of the system
always decrease. This is in contrast to the {\em equilibrium\/} energy and
angular momentum, which both increase for $r<r_m$. This point is illustrated
in Figure~6, where we show both the total energy of the system
during the evolution $\ce(r)$ and the equilibrium energy $E_{eq}(r)$.
At large $r$, $\ce \simeq E_{eq}$, but when $r\lo r_m$, $\ce<E_{eq}$.

The rapid coalescence of the binary for $r<r_m$ can affect
very significantly the gravitational
radiation waveforms. In Figure~7, we show the wave amplitude $h_{+}$
seen by an observer along the rotation axis ($\theta=0$).
In this case, $h_{\times}$ differs from $h_{+}$ only by a constant phase
of $\pi/2$ (cf. eqs.~[\ref{hplus}]-[\ref{hcross}]).
Here we show the irrotational ($\cc=0$) and corotating evolutions, and
compare them with the point-mass result,
for both $n=0.5$ and $n=1$, all with $R_o/M=5$.
We have set $t=0$ at $r=5R_o$, and the various solutions all
have the same phase at that point. We see that the phase and amplitude
depart significantly from the point mass result, even when $\cc=0$.
The departures are largest for the corotating case (but recall that the phase
error accumulates mainly at large $r$ in that case; cf.\ \S 3.2.4).
Figure~8 shows the number of cycles $N_{GW}=2 N_{orb}$
of the gravitational wave as a function of wave frequency during the evolution.
The wave frequency in real units is given by
\begin{equation}
f_{GW}=\frac{\Omega}{\pi}=4130\, M_{1.4}^{-1}\left({R_o\over 5M}\right)^{-3/2}
\frac{\Omega}{(M/R_o^3)^{1/2}}\,{\rm Hz},
\end{equation}
where $\Omega$ is given by equation (\ref{Omdef}) (and remains only
approximately
equal to the equilibrium value, eq.~[\ref{kepler}]).
We also see from Figures~6 and~7 that, for a given
$R_o/M$, finite-size effects on the waveforms are more important for
smaller $n$ (stiffer EOS).

In Table~4, we summarize our results for different cases.
The first two blocks in the table correspond to constant-$\cc$ solutions
with $\hat\Omega_s=0$ and $\hat \Omega_s=0.4$. The last block corresponds to
corotating evolutions. For every binary model, we list values of
the stability limit $r_m$, and the separation at contact, $r_f$.
In each case, we give the key orbital evolution parameters when
$R_o/M=5$ and $R_o/M=8$: the radial velocity at $r=r_m$, the radial velocity
at $r=r_f$, the maximum gravitational wave frequency
$f_{GW}^{(max)}=f_{GW}(r=r_f)$,
the maximum wave amplitude $h_{max}$, and the number of cycles
$N_{GW}$ of gravitational radiation from $r=5R_o$ to $r=r_f$.

\section{GENERAL RELATIVISTIC EFFECTS}

Our treatment so far has ignored Post-Newtonian (PN) effects other than
the lowest-order dissipative effect corresponding to the emission of
gravitational radiation according to the quadrupole formula~(\ref{quadfor}).
However, for the typical value of $R_o/M=5$, other PN effects are likely to be
important and can alter the orbits considerably.

In the case of two {\em point masses\/},
these PN effects can make a circular orbit become unstable when
the separation is smaller than some critical value
(``inner-most stable orbit'')
$r_{GR}$. Kidder, Will \& Wiseman (1992)  have recently obtained
\begin{equation}
r_{GR} \simeq 6M_t+4\mu,  \label{rGRdef}
\end{equation}
where $M_t=M+M'$ is the total mass\footnote{Their original result
was derived in harmonic coordinates. Here we ignore coordinate distinctions
and fit their result to equation (\ref{rGRdef}) as an approximation.}.
Their method includes PN effects up to order $(v/c)^4$
in the treatment of the two-body interaction, but also incorporates
test-particle effects in the Schwarzschild geometry exactly.
Indeed, for a test particle orbiting a spherical black hole,
expression~(\ref{rGRdef}) reduces to the familiar Schwarzschild result
$r_{GR}=6M_t$. For two point masses with $M=M'$,
expression~(\ref{rGRdef}) gives
the result\footnote{Note that this is very different
from the early estimate of Clark \& Eardley (1977), who give
$r_{GR} \simeq 6M$.} $r_{GR} \simeq 14 M$.
Since $r_m \simeq 3 R_o$ typically, we see that
$r_{GR}$ and $r_m$ are comparable for $R_o/M=5$.

For a test particle orbiting a spherical black hole of mass $M_t$,
the total energy describing the orbital motion is given by
\begin{equation}
\left(\frac{E}{\mu}\right)^2=\left(1-\frac{2M_t}{r}\right)
\left(1+\frac{J^2}{\mu^2 r^2}\right),   \label{Etp}
\end{equation}
where $r$ is the Schwarzschild radial coordinate.
The corresponding equilibrium energy for stationary circular orbits is
\begin{equation}
\left(\frac{E_{eq}}{\mu}\right)^2 = \frac{(r-2M_t)^2}{r(r-3M_t)}, \label{Eeqtp}
\end{equation}
which has a minimum at $r=r_{GR}= 6M_t$.
Although equation~(\ref{Eeqtp}) is very different from our purely Newtonian
expression~(\ref{Eeqdef}), the two have one thing in common,
namely the existence of a minimum marking the onset of instability.
Thus we see that Newtonian tidal effects and relativistic effects both
lead to the same qualitative result: the existence of a critical
binary separation where a circular orbit becomes dynamically unstable.

For two stars with comparable masses, a simple analytic expression such
as~(\ref{Eeqtp}) cannot be written since stationary circular orbits do
not exist for a system radiating gravitational waves. Kidder, Will, \&
Wiseman avoided this problem by artificially turning off the radiation reaction
terms in their PN equations of motion. In the same spirit, we adopt the
following simple ansatz for the energy of the point-mass binary system
\begin{equation}
\left(\frac{E^{(GR)}}{\mu}\right)^2=1-\frac{2M_t}{r}+\frac{J^2}{\mu^2r^2}
-\frac{2M_t+4\mu/3}{r^3}\left(\frac{J}{\mu}\right)^2.  \label{EGR}
\end{equation}
The corresponding equilibrium energy for circular orbits is obtained
by solving $(\partial E^{(GR)}/\partial r)_{J,M,M'}=0$, which gives
\begin{equation}
\left(\frac{E_{eq}^{(GR)}}{\mu}\right)^2
=\frac{(r-2M_t)^2-(4\mu/3)(3r/2-2M_t)}{r(r-3M_t-2\mu)}, \label{EGReq}
\end{equation}
and has a minimum at $r_{GR}$ given by equation~(\ref{rGRdef}).
In equation (\ref{EGR}),
the coefficient $4\mu/3$ was chosen to obtain this result. Note that
when $\mu\rightarrow 0$, equations~(\ref{EGR}) and~(\ref{EGReq}) reduce
to the exact results for a test particle, equations~(\ref{Etp})
and~(\ref{Eeqtp}).

To estimate the combined effects of general relativity and the Newtonian
tidal interactions for finite-size stars,
we adopt the simple model introduced in Paper~I.
The equilibrium energy of the Newtonian fluid system discussed in \S 2
can be modeled approximately as
\begin{equation}
E_{eq}^{(N)} = -\frac{M M'}{2 r}
+\frac{1}{2 \alpha}\frac{M M' r_m^{\alpha-1}}{r^{\alpha}},  \label{ENeq}
\end{equation}
which has a minimum at $r=r_m$. The parameters $\alpha$ and $r_m$
depend on the internal structure of the stars and the degree of
synchronization. Their values are adjusted to obtain the best possible
fit to the more accurate $E_{eq}$ calculated in \S2.
The total energy of the binary, not necessarily in equilibrium, is written as
\begin{equation}
\ce^{(N)}=\frac{\mu}{2}\left(\frac{dr}{dt}\right)^2+\frac{J^2}{2 \mu r^2}
-\frac{MM'}{r}-\frac{MM'r_m^{\alpha-1}}{2 \alpha (\alpha-1) r^{\alpha}}.
\label{ceN}
\end{equation}
For a stationary orbit (${\dot r}=0$), the equilibrium condition
$(\partial E/\partial r)_{J,M,M'}=0$ yields the equilibrium energy
given by equation~(\ref{ENeq}). Taking the time derivative of
equation~(\ref{ceN}), we obtain the orbital evolution equation
\begin{equation}
\mu{\ddot r}-\frac{J^2}{\mu r^3}+
+\frac{MM'}{r^2}+\frac{MM'r_m^{\alpha-1}}{(\alpha-2)r^{\alpha+1}}=0.
\label{rddotS}
\end{equation}
This equation, together with ${\dot J}={\dot E}/\Omega=
(\mu r^2/J){\dot E}$, can be solved for $r(t)$.
Equivalently, one can also eliminate $J$ and derive an equation
similar to~(\ref{rdddot}),
\begin{equation}
\mu \frac{d^3r}{dt^3} + \frac{3 \mu}{r}\left(\frac{dr}{dt}\right)
\left(\frac{d^2r}{dt^2}\right)
+\frac{2}{r}\frac{dE_{eq}}{dr}\frac{dr}{dt}
=\frac{2}{r} \left(\frac{d\ce}{dt}\right)_{GW}.   \label{rdddotS}
\end{equation}

The solutions of equation~(\ref{rdddotS}) reproduce all the essential features
of the more accurate solutions obtained in \S 4.
For example, for a typical case with $r_m=2.8R_o$,
$r_f=2.5R_o$ and $\alpha=6$, we get $\dot r(r_m)=v_r(r_m)=-0.059(M/R_o)^{1/2}$
and $v_r(r_f)=-0.12(M/R_o)^{1/2}$ for $R_o/M=5$. These are close to the
the typical values found in \S 4.

We now attempt to incorporate the new relativistic effects discussed above into
the model. We simply replace the Newtonian point-mass term $-MM'/(2r)$ in
equation~(\ref{ENeq}) by the corresponding GR term, expression~(\ref{EGReq}),
\begin{equation}
E_{eq}=E_{eq}^{(N)}+\frac{MM'}{2r}+E_{eq}^{(GR)}-\mu
=\frac{1}{2 \alpha}\frac{M M' r_m^{\alpha-1}}{r^{\alpha}}
+E_{eq}^{(GR)}-\mu.   \label{ENGReq}
\end{equation}
Of course, relativistic effects will also change the internal
structures of the stars, changing the first term in
the first equality of (\ref{ENGReq}), but this is a higher order effect.
The new $E_{eq}(r)$ now has a minimum at some $r_m'$ which is
larger than either the Newtonian $r_m$ or the purely relativistic
$r_{GR}$. For example, for $r_m=2.8R_o$ and $R_o/M=5$, we find
$r_m'=3.4R_o$. To calculate the orbital evolution,
we use the new $E_{eq}(r)$ in equation~(\ref{rdddotS})
in place of $E_{eq}^{(N)}$. In the example considered above, we obtain
$v_r(r_m')=-0.048(M/R_o)^{1/2}$ and $v_r(r_f)=-0.25(M/R_o)^{1/2}$, i.e.,
the terminal velocity at contact is about a factor of two larger than
when only Newtonian effects are considered.
For $R_o/M=8$, we have $r_m'=3.0R_o$, and the terminal velocities are
$v_r(r_m')=-0.028(M/R_o)^{1/2}$ and $v_r(r_f)=-0.13(M/R_o)^{1/2}$.
We see that the general relativistic effects can be important,
especially when the value of $R_o/M$ is small.

Our treatment of general relativistic effects is admittedly very crude.
The main point we wish to emphasize here is that
the Newtonian hydrodynamic effects discussed in this paper
are likely to be at least as important as the relativistic corrections
to the orbital motion for the final coalescence of neutron star binaries.
When the two effects are combined, the final coalescence is likely to be
even faster, and may assume a significant ``head-on'' character.

\section{POSSIBILITY OF MASS TRANSFER}

Our discussion so far has assumed that there is no mass loss from
the system or mass transfer between the two stars.
For nearly-equal-mass binaries, equilibrium configurations exists
all the way down to contact (and even beyond; see Hachisu 1986),
and there is no ``Roche limit'' in the conventional sense.
When the two masses are sufficiently different, however, a Roche limit
can exist and mass transfer becomes a possibility (see \S 2.4, 2.5).
In particular, Clark \& Eardley (1977) have
suggested that stable mass transfer from the less massive neutron star to
the more massive one can occur at the Roche limit. This stable mass
transfer phase may last hundreds of orbital revolutions before the
lighter star is tidally disrupted. It is accompanied by a secular {\em
increase\/}
of the orbital separation. Thus if stable mass transfer indeed occurs,
a characteristic  ``reversed chirp'' would be observed in
the gravitational wave signal (Jaranowski \& Krolak 1992).
The problem has been reexamined more recently by Kochanek (1992)
and Bildsten \& Cutler (1992), who both find that very large mass transfer
rates and extreme mass ratios are required for stable mass transfer
between two neutron stars, making it rather unlikely.

Our results suggest an additional problem with the Clark-Eardley scenario.
Quite independent from the stability of the mass transfer itself,
the Clark-Eardley scenario requires the existence of a dynamically stable
Roche limit configuration.
However, our studies of Roche-Riemann and Darwin-Riemann equilibrium
configurations (LRS1, LRS2) indicate that this is almost impossible for
objects as incompressible as neutron stars. Consider again the equilibrium
energy curve $E_{eq}(r)$ for a binary. The Roche limit at
$r=r_{lim}$, when it exists, corresponds to the point where
$r$ has a minimum possible value for circular equilibrium prior to contact.
However, before such a limit can be reached, the binary separation
must pass through a value $r_m>r_{lim}$ where $E_{eq}(r)$ is minimum.
For equilibrium  sequences with constant $\cc$, this minimum  coincides with
the dynamical stability limit, and all binaries with $r<r_m$ are
dynamically unstable (cf.\ \S 2.4).
Therefore, if viscosity can be neglected in the
neutron star binary evolution, no dynamically stable Roche limit can
exist, and {\em stable mass transfer can never occur}.

In the opposite limit, when viscosity is so efficient that
corotation can be maintained throughout the evolution,
the mass parameter range which permits the existence of a stable Roche limit
is very small. This is illustrated in Figure~9, which shows the different
types of terminal binary neutron star configurations as a function of $M$
and $M'$. The neutron star model is
based on the WFF equation of state AV14+UVII, with the effective
polytropic index given in Table~3 (see \S 4.1). The diagrams are constructed
from our general Darwin-Riemann equilibrium models allowing for two
nonidentical
polytropes (see LRS2 for details and for applications to other types of binary
systems).
For irrotational ($\cc=0$) systems (Fig.~9a), the existence of a Roche limit
requires
a mass ratio $q\lo 0.9$ or $q \go 1.1$ for $M \simeq 1.4 M_{\odot}$.
However, as discussed above, this Roche limit is always dynamically unstable.
For corotating systems (Fig.~9b), the existence of a Roche limit
requires a mass ratio $q\lo 0.8$ or $q\go 1.2$
for $M\simeq 1.4M_\odot$. Stable configurations can exist at the
Roche limit but only if one of the stars has a very small mass,
$M\lo 0.4 M_{\odot}$.

We conclude that even when the masses of the two neutron stars
are different, stable mass transfer is nearly impossible.
The final phase of the orbital decay is always a rapid coalescence,
and a ``reverse chirp'' in the gravitational wave signal is not expected.

\clearpage

\acknowledgments

This work has been supported by NSF Grant AST 91--19475 and NASA Grant
NAGW--2364
to Cornell University, and by a Hubble Fellowship to F.~A.~R. funded by NASA
through Grant HF-1037.01-92A
awarded by the Space Telescope Science Institute, which is operated by the
Association of Universities for Research in Astronomy, Inc.,
for NASA, under contract NAS5-26555.

\clearpage

\begin{figure}
\caption{
Equilibrium curves of total energy, total angular momentum and orbital
angular velocity as a function of binary separation along various
sequences with $n=0.5$. Here $\Omega_k=(2M/r^3)^{1/2}$ is the Keplerian
angular velocity.
Solid line: irrotational ($\cc=0$) Darwin-Riemann sequence;
dotted line: Darwin-Riemann sequence with $\cc=-0.0652$, corresponding
to an initial spin ${\hat \Omega_s}=\Omega_s/(M/R_o^3)^{1/2}=0.1$ for
both stars; short dashed line: $\cc=-0.1304$
(${\hat \Omega_s}=0.2$); long dashed line: $\cc=-0.2607$
(${\hat\Omega_s}=0.4$); light dotted-dashed line:
corotating (Darwin) sequence ($f_R=0$).
}
\end{figure}

\begin{figure}
\caption{
Variation of the three principal axes $a_1,~a_2$ and $a_3$
along the same sequences shown in Fig.~1.
}
\end{figure}

\begin{figure}
\caption{
The accumulated change in the number of cycles of gravitational wave
due to finite-size effects at large separation. The change is shown
as a function of wave frequency $f_{GW}$ (starting with zero
change at $f_{GW}=10\,$Hz).
The two stars are assumed to be identical, with $M=1.4M_{\odot}$,
$R_o/M=5$, and $n=0.5$.
The dotted line shows the tidal effects in irrotational
configurations, $\delta N^{(I)}_{GW}$ (eq.~[90]);
the dashed lines show the spin-induced change $\delta N^{(S)}_{GW}$
(eq.~[91]), for $\hat{\Omega_s}=0.1$ and $\hat{\Omega_s}=0.2$;
the solid lines show the combined tidal and spin-induced effects
($\delta N^{(I)}_{GW}+\delta N^{(S)}_{GW}$).
}
\end{figure}

\begin{figure}
\caption{
Terminal evolution of selected binary models shown in Fig.~1. Here
(a) shows the infall radial
velocity $v_r=\dot r$, (b) shows the time (contact is reached at $t=0$)
and (c) shows the number of orbits, starting from $r=5R_o$.
In all cases the two stars are identical with $n=0.5,~R_o/M=5$.
The solid line is for $\hat \Omega_s=0~(\cc=0)$, the
short-dashed line for $\hat \Omega_s=0.2$, the long-dashed line
for $\hat \Omega_s=0.4$, and the dotted-dashed line for corotating evolution.
All curves terminate when contact is reached. For comparison,
the dotted lines show the results for two point masses.
}
\end{figure}

\begin{figure}
\caption{
Terminal evolution of selected binary models with different $n$.
In all cases the two stars are identical and have $R_o/M=5$.
Here (a) shows the infall radial velocity and (b) shows the number of
orbits starting from $r=5R_o$. The solid line is for $\cc=0,~n=0.5$,
the short-dashed line for $\cc=0,~n=1$, the long-dashed line
for corotation and $n=0.5$, and the dotted-dashed
line for corotation and $n=1$. The dotted line is the point-mass result.
}
\end{figure}

\begin{figure}
\caption{
Total energy of the binary system as $r$ decreases with time
for identical configurations with
$n=0.5$ and $R_o/M=5$. The solid line is for $\hat \Omega_s=0$,
the short-dashed line for $\hat \Omega_s=0.2$, and the long-dashed line
for corotation. The dotted lines are the corresponding equilibrium energy
curves (as shown in Fig.~1).
}
\end{figure}

\begin{figure}
\caption{
Gravitational radiation waveform $h_{+}$ just prior to contact.
Here (a) is for $R_o/M=5,~n=1$, and (b) for
$R_o/M=5,~n=0.5$. The solid lines are for irrotational
Darwin-Riemann binaries ($\cc=0$), the dashed lines for corotating
(Darwin) binaries,
and the dotted lines for two point masses; the long-dashed lines
show the envelopes of the waveforms. $D$ is the distance from
the binary to the observer located along the binary axis ($\theta=0$).
}
\end{figure}

\begin{figure}
\caption{
Number of gravitational wave cycles $N_{GW}=2N_{orb}$
as a function of the wave frequency $f_{GW}$ just prior to contact.
Here $R_o/M=5$ and  $M=1.4M_{\odot}$.
The solid line is for an irrotational binary with $n=0.5$,
the short-dashed line for an irrotational binary with $n=1$,
the long-dashed line for a corotating binary with $n=0.5$,
and the dotted-dashed line for a corotating binary with $n=1$.
The light dotted line shows the point-mass result for comparison.
}
\end{figure}

\begin{figure}
\caption{
Final fates of coalescing neutron star binaries.
Each point in the diagram corresponds
to a sequence of binary equilibrium configurations with different values of $M$
and $M'$.
The various regions indicate the possible terminal configurations.
Neutron star models are based on the WFF equation of state.
Irrotational systems are shown in (a), corotating systems in (b).
In (a), no binary equilibrium sequence in
the hatched regions has a Roche limit. Instead, a contact
equilibrium configuration exists. In the wide-hatched region,
this contact configuration is dynamically unstable,
while it is dynamically stable in the narrow-hatched region.
A binary sequence in the unshaded region does terminate at a  Roche limit,
but the Roche limit configuration is always dynamically unstable.
In (b), no
binary sequence in the shaded regions inside the solid lines has a Roche limit.
The contact configuration is dynamically unstable in the wide-hatched region,
while it is dynamically stable in the narrow-hatched region.
Binary sequences outside the solid lines have Roche limits.
A binary in the unshaded
region encounters a dynamical instability before the Roche limit; this
instability occurs after the Roche limit (along the second, unphysical branch
of the equilibrium sequence, cf.\ \S2.4) in the cross-hatched region.
A binary in the blackened region encounters a Roche limit but
no dynamical instability. Only in the cross-hatched
and blackened regions can a dynamically stable binary exist at the Roche limit.
Binaries in all regions in (b) encounter a secular instability (energy minimum)
prior to any other critical point, except those in
the small unshaded portion in the lower left corner.
}
\end{figure}

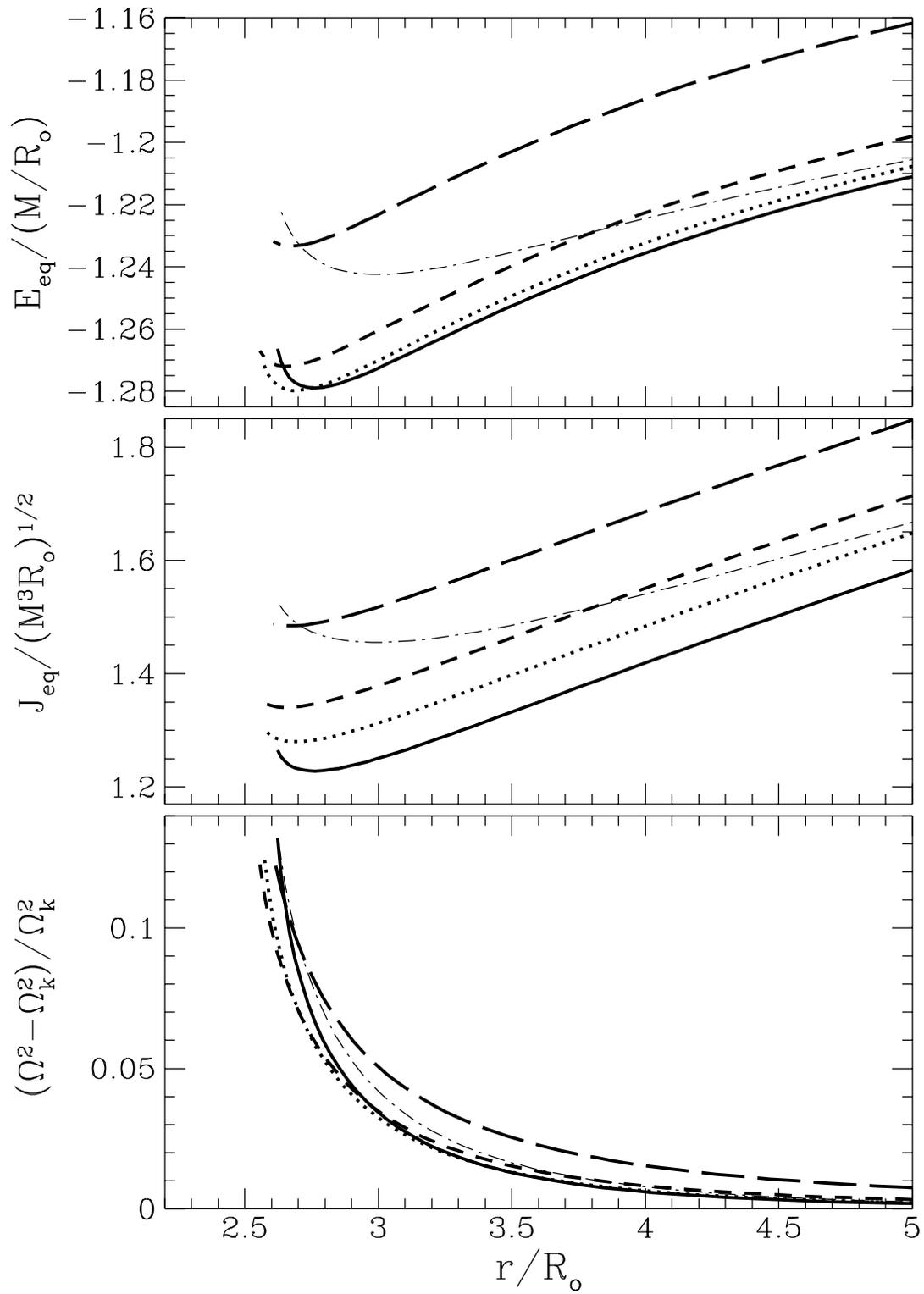
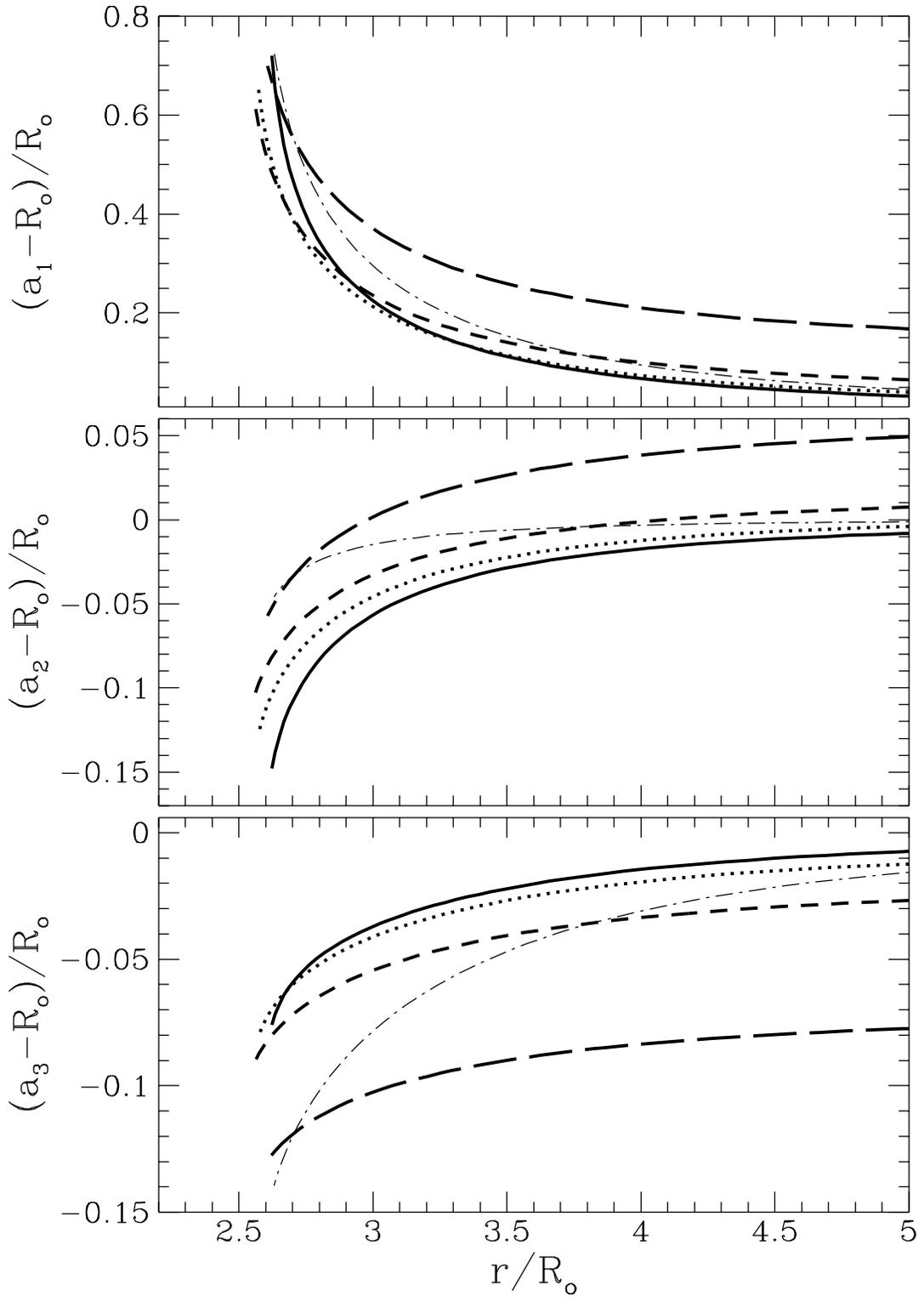
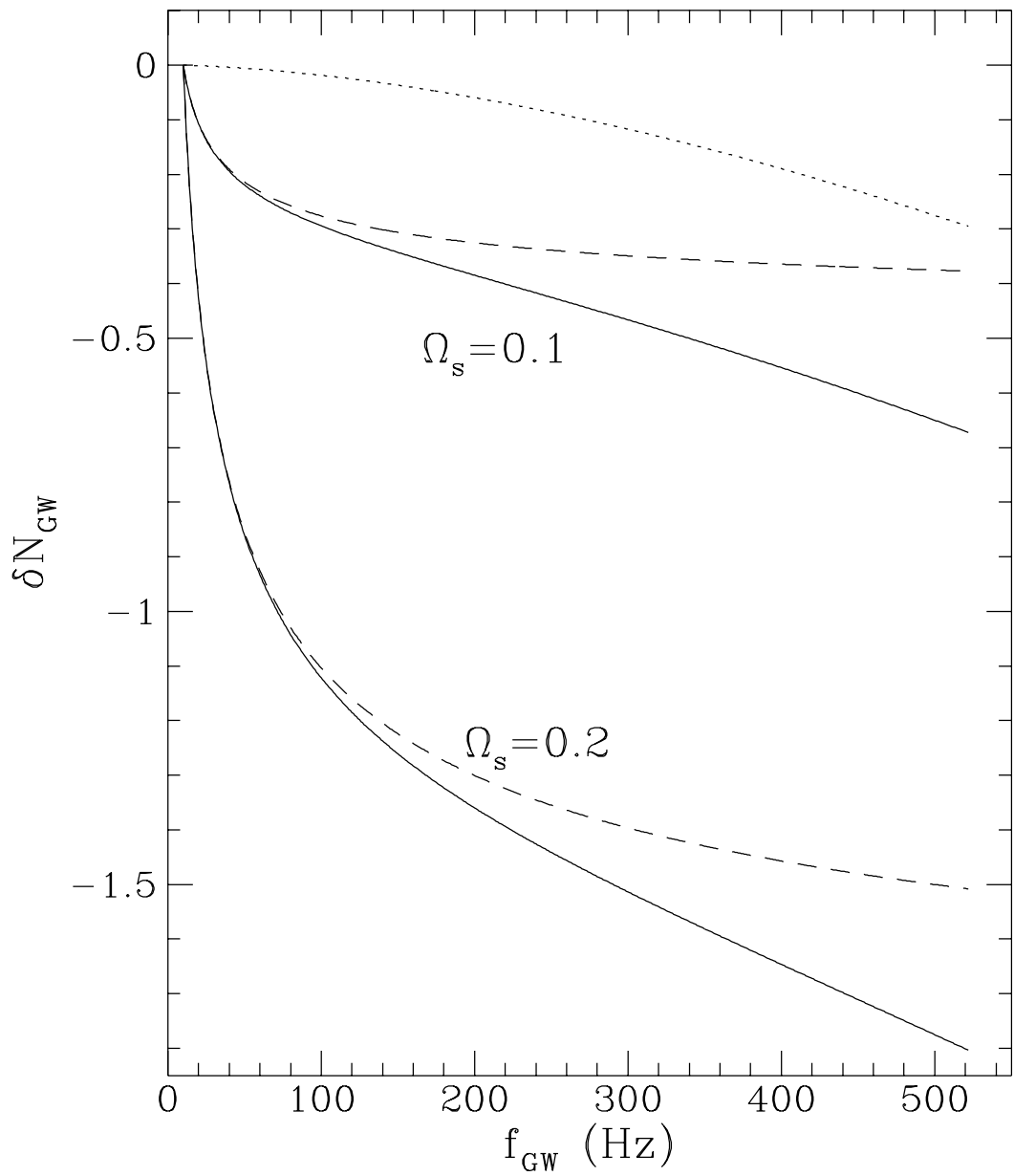
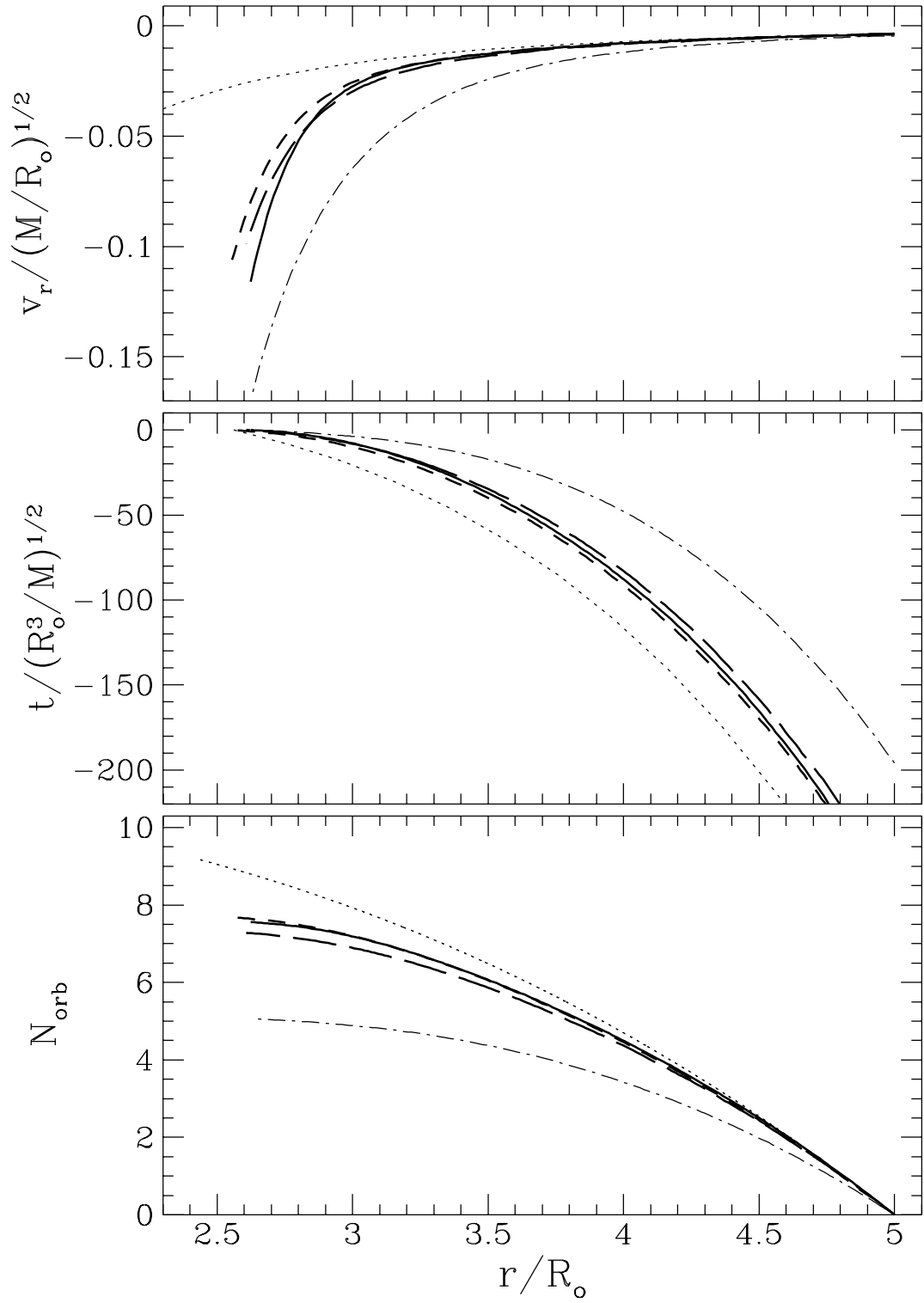
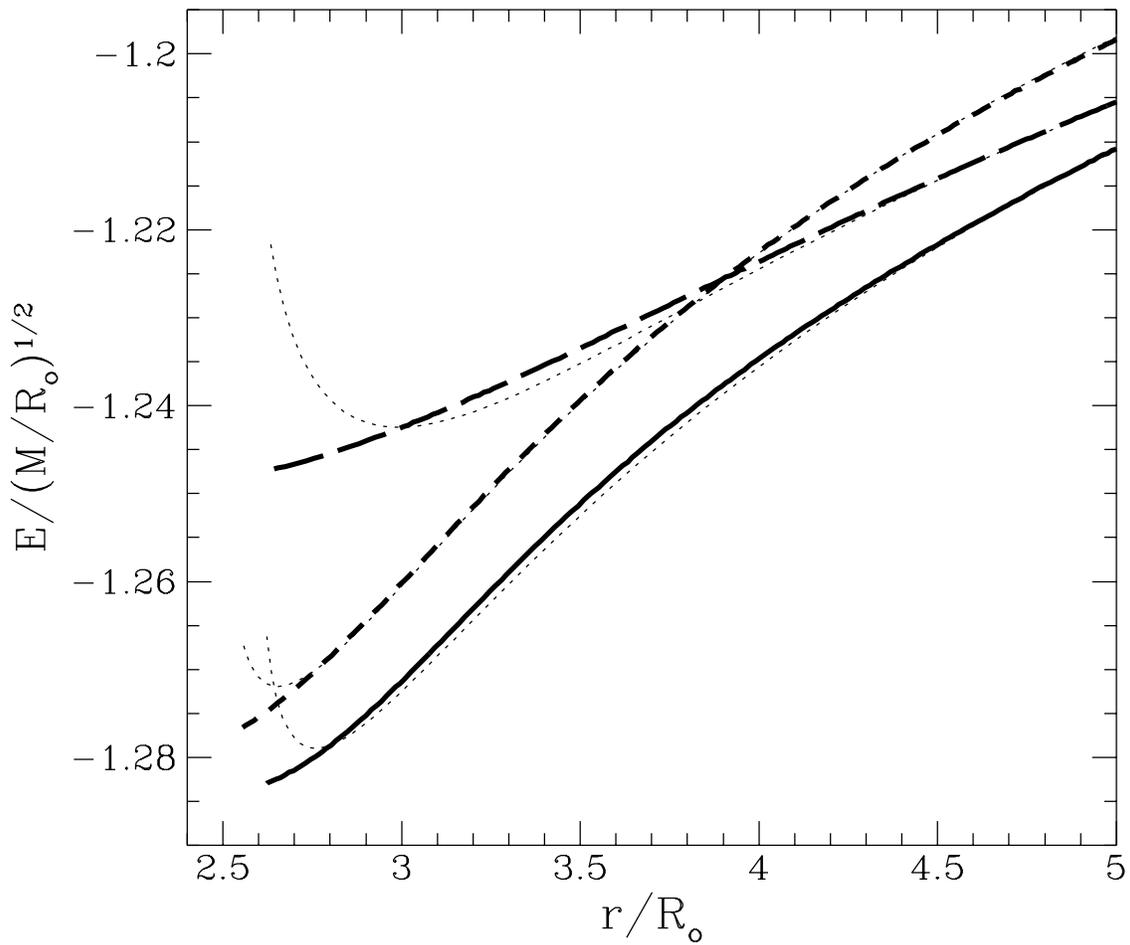
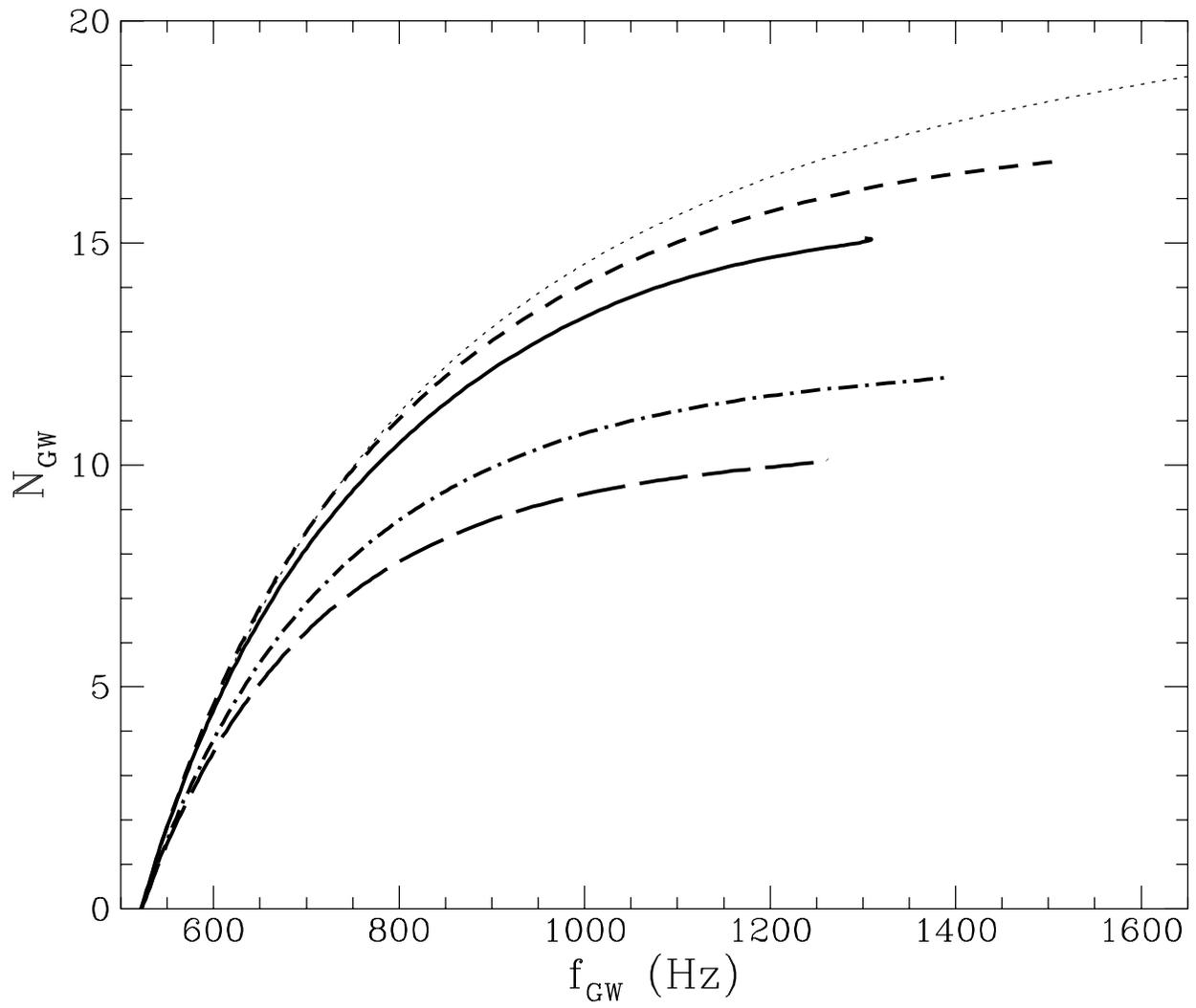
\end{document}


\newpage
\clearpage
\begin{table}
\begin{center}
TABLE 1\\
{\sc Compressible Darwin-Riemann Sequences With ${\cal C}=0\,\,{}^a$} \\
\vspace {.1in}
\small
\begin{tabular}{l l l l l l l l l}
\hline
\hline
\multicolumn{9}{l}{\vspace{-.12in}}\\
$r/a_1~^b$ & $r/R_0$ & $a_2/a_1$ & $a_3/a_1$ & $T_s/|W|$ & $~~\bar\Omega$ 
&$~~\bar J$& $~~~\bar E$ & $R/R_0$\\
\multicolumn{9}{l}{\vspace{-.12in}}\\
\hline
\multicolumn{9}{l}{\vspace{-.12in}}\\
\multicolumn{9}{c}{$n=0$}\\
\multicolumn{9}{l}{\vspace{-.12in}}\\
\hline
\multicolumn{9}{l}{\vspace{-.12in}}\\
 6.0  & 6.071& 0.9824& 0.9828& 0.945(-6)& 0.1092&  1.743&-1.2823& 1.\\
 5.0  & 5.103& 0.9693& 0.9705& 0.492(-5)& 0.1419&  1.600&-1.2978& 1.\\
 4.5  & 4.628& 0.9576& 0.9597& 0.128(-4)& 0.1644&  1.525&-1.3076& 1.\\
 4.0  & 4.165& 0.9391& 0.9430& 0.374(-4)& 0.1929&  1.450&-1.3193& 1.\\
 3.5  & 3.722& 0.9078& 0.9160& 0.126(-3)& 0.2293&  1.377&-1.3325& 1.\\
 3.2  & 3.472& 0.8783& 0.8912& 0.284(-3)& 0.2556&  1.337&-1.3408& 1.\\
 3.0  & 3.317& 0.8514& 0.8692& 0.508(-3)& 0.2750&  1.315&-1.3460& 1.\\
 2.8  & 3.173& 0.8167& 0.8411& 0.937(-3)& 0.2957&  1.298&-1.3501& 1.\\
 2.582*& 3.037& 0.7671& 0.8013& 0.189(-2)& 0.3191&  1.291&-1.3521& 1.\\
 2.4  & 2.944& 0.7142& 0.7583& 0.345(-2)& 0.3383&  1.298&-1.3500& 1.\\
 2.2  & 2.873& 0.6429& 0.6986& 0.673(-2)& 0.3575&  1.329&-1.3406& 1.\\
 2.0  & 2.842& 0.5586& 0.6238& 0.130(-1)& 0.3725&  1.397&-1.3190& 1.\\
\multicolumn{9}{l}{\vspace{-.12in}}\\
\hline
\multicolumn{9}{l}{\vspace{-.12in}}\\
\multicolumn{9}{c}{$n=0.5$}\\
\multicolumn{9}{l}{\vspace{-.12in}}\\
\hline
\multicolumn{9}{l}{\vspace{-.12in}}\\
 6.0  & 6.052& 0.9871& 0.9874& 0.371(-6)& 0.1097&  1.740&-1.1937& 1.0000\\
 5.0  & 5.075& 0.9776& 0.9783& 0.193(-5)& 0.1430&  1.594&-1.2095& 1.0000\\
 4.0  & 4.120& 0.9557& 0.9579& 0.146(-4)& 0.1958&  1.439&-1.2320& 1.0002\\
 3.5  & 3.661& 0.9331& 0.9378& 0.492(-4)& 0.2343&  1.360&-1.2467& 1.0004\\
 3.2  & 3.396& 0.9117& 0.9193& 0.111(-3)& 0.2629&  1.314&-1.2565& 1.0006\\
 3.0  & 3.228& 0.8921& 0.9027& 0.200(-3)& 0.2846&  1.286&-1.2633& 1.0009\\
 2.8  & 3.068& 0.8664& 0.8815& 0.373(-3)& 0.3084&  1.260&-1.2698& 1.0014\\
 2.6  & 2.920& 0.8322& 0.8537& 0.724(-3)& 0.3340&  1.240&-1.2754& 1.0023\\
 2.5  & 2.853& 0.8111& 0.8367& 0.102(-2)& 0.3473&  1.233&-1.2775& 1.0029\\
 2.344*& 2.759& 0.7716& 0.8049& 0.178(-2)& 0.3681&  1.228&-1.2790& 1.0042\\
 2.2  & 2.688& 0.7267& 0.7685& 0.302(-2)& 0.3867&  1.234&-1.2772& 1.0062\\
 2.0  & 2.623& 0.6498& 0.7045& 0.635(-2)& 0.4091&  1.265&-1.2663& 1.0107\\
\multicolumn{9}{l}{\vspace{-.12in}}\\
\hline
\multicolumn{9}{l}{\vspace{-.12in}}\\
\multicolumn{9}{c}{$n=1.0$}\\
\multicolumn{9}{l}{\vspace{-.12in}}\\
\hline
\multicolumn{9}{l}{\vspace{-.12in}}\\
 6.0  & 6.037& 0.9909& 0.9910& 0.134(-6)& 0.1101&  1.738&-1.0828& 1.0000\\
 5.0  & 5.053& 0.9841& 0.9844& 0.695(-6)& 0.1438&  1.590&-1.0989& 1.0001\\
 4.0  & 4.085& 0.9687& 0.9698& 0.524(-5)& 0.1981&  1.431&-1.1221& 1.0002\\
 3.5  & 3.613& 0.9528& 0.9553& 0.176(-4)& 0.2384&  1.348&-1.1378& 1.0005\\
 3.2  & 3.338& 0.9378& 0.9419& 0.399(-4)& 0.2690&  1.298&-1.1488& 1.0008\\
 3.0  & 3.159& 0.9240& 0.9298& 0.717(-4)& 0.2926&  1.265&-1.1567& 1.0012\\
 2.8  & 2.987& 0.9058& 0.9143& 0.134(-3)& 0.3190&  1.234&-1.1650& 1.0018\\
 2.6  & 2.823& 0.8814& 0.8939& 0.263(-3)& 0.3485&  1.205&-1.1733& 1.0029\\
 2.4  & 2.672& 0.8482& 0.8666& 0.541(-3)& 0.3806&  1.181&-1.1809& 1.0047\\
 2.2  & 2.539& 0.8024& 0.8297& 0.117(-2)& 0.4143&  1.167&-1.1859& 1.0079\\
 2.115*& 2.491& 0.7779& 0.8100& 0.164(-2)& 0.4285&  1.165&-1.1866& 1.0100\\
 2.0  & 2.437& 0.7396& 0.7790& 0.262(-2)& 0.4466&  1.169&-1.1851& 1.0140\\
\multicolumn{9}{l}{\vspace{-.12in}}\\
\hline
\multicolumn{9}{l}{\vspace{-.12in}}\\
\multicolumn{9}{c}{$n=1.5$}\\
\multicolumn{9}{l}{\vspace{-.12in}}\\
\hline
\multicolumn{9}{l}{\vspace{-.12in}}\\
 6.0  & 6.025& 0.9938& 0.9938& 0.429(-7)& 0.1104&  1.736& -.9401& 1.0000\\
 5.0  & 5.036& 0.9892& 0.9893& 0.222(-6)& 0.1445&  1.587& -.9564& 1.0001\\
 4.0  & 4.058& 0.9787& 0.9793& 0.167(-5)& 0.1999&  1.426& -.9802& 1.0002\\
 3.5  & 3.577& 0.9680& 0.9692& 0.561(-5)& 0.2418&  1.339& -.9966& 1.0004\\
 3.2  & 3.293& 0.9578& 0.9599& 0.126(-4)& 0.2739&  1.287&-1.0084& 1.0008\\
 3.0  & 3.108& 0.9486& 0.9515& 0.227(-4)& 0.2991&  1.251&-1.0172& 1.0011\\
 2.8  & 2.926& 0.9363& 0.9406& 0.426(-4)& 0.3278&  1.216&-1.0268& 1.0017\\
 2.6  & 2.750& 0.9199& 0.9263& 0.836(-4)& 0.3605&  1.181&-1.0370& 1.0027\\
 2.4  & 2.582& 0.8973& 0.9071& 0.173(-3)& 0.3976&  1.149&-1.0476& 1.0043\\
 2.2  & 2.426& 0.8654& 0.8807& 0.380(-3)& 0.4386&  1.121&-1.0576& 1.0073\\
 2.1  & 2.355& 0.8447& 0.8638& 0.577(-3)& 0.4602&  1.110&-1.0619& 1.0098\\
 2.0  & 2.291& 0.8199& 0.8437& 0.890(-3)& 0.4819&  1.102&-1.0651& 1.0132\\
\multicolumn{9}{l}{\vspace{-.12in}}\\
\hline
\multicolumn{9}{l}{\vspace{-.1in}}\\
\multicolumn{9}{l}{$^a$ Here 
$\bar \Omega$, $\bar J$ and $\bar E$ are defined 
in eqs.~(34); $R$ and $R_o$ are defined by eqs.~(26)--(27);}\\
\multicolumn{9}{l}{$T_s$ is the ``spin'' kinetic energy (cf. eq.~[14])
and $W$ is given in eq.~(3).}\\
\multicolumn{9}{l}{$^b$ * marks the dynamical stability limit.}\\
\end{tabular}
\end{center}
\end{table}

\newpage
\clearpage
\begin{table}
\begin{center}
TABLE 2\\
{\sc Compressible Darwin-Riemann Sequences With Constant ${\cal C}\,\,{}^a$} \\
\vspace {.1in}
\small
\begin{tabular}{l l l l l l l l l}
\hline
\hline
\multicolumn{9}{l}{\vspace{-.12in}}\\
$r/a_1~^b$ & $r/R_0$ & $a_2/a_1$ & $a_3/a_1$ & $T_s/|W|$ & $~~\bar\Omega$ 
&$~~\bar J$& $~~~\bar E$ & $R/R_0$\\
\multicolumn{9}{l}{\vspace{-.12in}}\\
\hline
\multicolumn{9}{l}{\vspace{-.12in}}\\
\multicolumn{9}{c}{$n=0.5,~~~{\hat\Omega_s}=0.2~({\cal C}=-0.1304)$}\\
\multicolumn{9}{l}{\vspace{-.12in}}\\
\hline
\multicolumn{9}{l}{\vspace{-.12in}}\\
 6.0  & 6.146& 0.9873& 0.9532& 0.947(-2)& 0.1072&  1.885&-1.1796& 1.0039\\
 5.0  & 5.153& 0.9784& 0.9450& 0.945(-2)& 0.1398&  1.738&-1.1952& 1.0039\\
 4.5  & 4.662& 0.9706& 0.9379& 0.943(-2)& 0.1626&  1.660&-1.2053& 1.0040\\
 4.0  & 4.177& 0.9586& 0.9270& 0.940(-2)& 0.1919&  1.581&-1.2175& 1.0040\\
 3.5  & 3.705& 0.9389& 0.9095& 0.936(-2)& 0.2303&  1.500&-1.2322& 1.0042\\
 3.2  & 3.430& 0.9208& 0.8937& 0.935(-2)& 0.2592&  1.452&-1.2424& 1.0045\\
 3.0  & 3.253& 0.9045& 0.8796& 0.936(-2)& 0.2814&  1.421&-1.2496& 1.0047\\
 2.8  & 3.083& 0.8833& 0.8617& 0.941(-2)& 0.3062&  1.392&-1.2569& 1.0051\\
 2.6  & 2.921& 0.8555& 0.8384& 0.956(-2)& 0.3336&  1.367&-1.2639& 1.0058\\
 2.4  & 2.774& 0.8183& 0.8078& 0.993(-2)& 0.3634&  1.348&-1.2696& 1.0070\\
 2.210*& 2.654& 0.7709& 0.7691& 0.108(-1)& 0.3928&  1.341&-1.2719& 1.0088\\
 2.0  & 2.555& 0.7001& 0.7111& 0.130(-1)& 0.4236&  1.354&-1.2670& 1.0127\\
\multicolumn{9}{l}{\vspace{-.12in}}\\
\hline
\multicolumn{9}{l}{\vspace{-.12in}}\\
\multicolumn{9}{c}{$n=0.5,~~~{\hat\Omega_s}=0.4~({\cal C}=-0.2607)$}\\
\multicolumn{9}{l}{\vspace{-.12in}}\\
\hline
\multicolumn{9}{l}{\vspace{-.12in}}\\
 6.0  & 6.423& 0.9859& 0.8648& 0.352(-1)& 0.1005&  2.056&-1.1398& 1.0151\\
 5.0  & 5.384& 0.9766& 0.8578& 0.351(-1)& 0.1311&  1.906&-1.1546& 1.0152\\
 4.5  & 4.871& 0.9687& 0.8518& 0.350(-1)& 0.1525&  1.828&-1.1643& 1.0152\\
 4.0  & 4.363& 0.9567& 0.8428& 0.349(-1)& 0.1802&  1.746&-1.1760& 1.0153\\
 3.5  & 3.867& 0.9377& 0.8283& 0.348(-1)& 0.2166&  1.663&-1.1902& 1.0155\\
 3.2  & 3.577& 0.9206& 0.8153& 0.346(-1)& 0.2442&  1.614&-1.2001& 1.0157\\
 3.0  & 3.388& 0.9055& 0.8039& 0.345(-1)& 0.2656&  1.581&-1.2072& 1.0160\\
 2.8  & 3.206& 0.8862& 0.7894& 0.344(-1)& 0.2897&  1.551&-1.2145& 1.0164\\
 2.6  & 3.031& 0.8612& 0.7707& 0.342(-1)& 0.3168&  1.523&-1.2219& 1.0170\\
 2.4  & 2.868& 0.8282& 0.7461& 0.342(-1)& 0.3468&  1.500&-1.2285& 1.0180\\
 2.2  & 2.723& 0.7839& 0.7131& 0.344(-1)& 0.3791&  1.486&-1.2328& 1.0197\\
 2.120*& 2.672& 0.7620& 0.6968& 0.347(-1)& 0.3923&  1.484&-1.2334& 1.0207\\
 2.0  & 2.607& 0.7235& 0.6679& 0.354(-1)& 0.4117&  1.489&-1.2318& 1.0228\\
\multicolumn{9}{l}{\vspace{-.12in}}\\
\hline
\multicolumn{9}{l}{\vspace{-.12in}}\\
\multicolumn{9}{c}{$n=1,~~~{\hat\Omega_s}=0.2~({\cal C}=-0.1046)$}\\
\multicolumn{9}{l}{\vspace{-.12in}}\\
\hline
\multicolumn{9}{l}{\vspace{-.12in}}\\
 6.0  & 6.130& 0.9909& 0.9662& 0.679(-2)& 0.1077&  1.856&-1.0712& 1.0069\\
 5.0  & 5.130& 0.9845& 0.9602& 0.677(-2)& 0.1407&  1.708&-1.0871& 1.0070\\
 4.0  & 4.144& 0.9702& 0.9468& 0.674(-2)& 0.1939&  1.547&-1.1101& 1.0071\\
 3.5  & 3.662& 0.9559& 0.9337& 0.672(-2)& 0.2338&  1.462&-1.1257& 1.0074\\
 3.2  & 3.379& 0.9426& 0.9217& 0.671(-2)& 0.2642&  1.411&-1.1367& 1.0077\\
 3.0  & 3.195& 0.9306& 0.9111& 0.670(-2)& 0.2878&  1.377&-1.1448& 1.0080\\
 2.8  & 3.016& 0.9150& 0.8974& 0.671(-2)& 0.3146&  1.344&-1.1534& 1.0086\\
 2.6  & 2.843& 0.8944& 0.8797& 0.676(-2)& 0.3448&  1.313&-1.1623& 1.0095\\
 2.4  & 2.680& 0.8665& 0.8561& 0.689(-2)& 0.3786&  1.285&-1.1710& 1.0110\\
 2.2  & 2.533& 0.8282& 0.8242& 0.723(-2)& 0.4154&  1.264&-1.1783& 1.0137\\
 2.012*& 2.416& 0.7783& 0.7834& 0.806(-2)& 0.4509&  1.256&-1.1814& 1.0183\\
 2.0  & 2.410& 0.7746& 0.7803& 0.814(-2)& 0.4531&  1.256&-1.1814& 1.0187\\
\multicolumn{9}{l}{\vspace{-.12in}}\\
\hline
\multicolumn{9}{l}{\vspace{-.12in}}\\
\multicolumn{9}{c}{$n=1,~~~{\hat\Omega_s}=0.4~({\cal C}=-0.2091)$}\\
\multicolumn{9}{l}{\vspace{-.12in}}\\
\hline
\multicolumn{9}{l}{\vspace{-.12in}}\\
 6.0  & 6.401& 0.9903& 0.9007& 0.254(-1)& 0.1009&  2.000&-1.0386& 1.0270\\
 5.0  & 5.357& 0.9836& 0.8953& 0.254(-1)& 0.1319&  1.849&-1.0538& 1.0270\\
 4.0  & 4.327& 0.9693& 0.8836& 0.253(-1)& 0.1821&  1.685&-1.0758& 1.0272\\
 3.5  & 3.822& 0.9553& 0.8723& 0.252(-1)& 0.2197&  1.599&-1.0907& 1.0275\\
 3.2  & 3.525& 0.9426& 0.8620& 0.251(-1)& 0.2485&  1.546&-1.1014& 1.0278\\
 3.0  & 3.331& 0.9314& 0.8529& 0.250(-1)& 0.2710&  1.511&-1.1092& 1.0282\\
 2.8  & 3.141& 0.9169& 0.8413& 0.249(-1)& 0.2967&  1.477&-1.1177& 1.0287\\
 2.6  & 2.957& 0.8980& 0.8263& 0.248(-1)& 0.3259&  1.444&-1.1266& 1.0295\\
 2.4  & 2.781& 0.8728& 0.8065& 0.247(-1)& 0.3590&  1.413&-1.1355& 1.0309\\
 2.2  & 2.619& 0.8386& 0.7798& 0.248(-1)& 0.3959&  1.388&-1.1437& 1.0333\\
 2.0  & 2.478& 0.7909& 0.7430& 0.251(-1)& 0.4352&  1.374&-1.1490& 1.0377\\
\multicolumn{9}{l}{\vspace{-.12in}}\\
\hline
\multicolumn{9}{l}{\vspace{-.1in}}\\
\multicolumn{9}{l}{$^a$ Here  
$\bar \Omega$, $\bar J$ and $\bar E$ are defined 
in eq.~(34); $R$ and $R_o$ are defined by eqs.~(26)--(27);}\\
\multicolumn{9}{l}{$T_s$ is the ``spin'' kinetic energy (cf. eq.~[14])
and $W$ is given in eq.~(3).}\\
\multicolumn{9}{l}{$^b$ * marks the dynamical stability limit.}\\
\end{tabular}
\end{center}
\end{table}

\newpage
\clearpage
\begin{table}
\begin{center}
TABLE 3\\
{\sc Polytropic Model for Neutron Stars$\,\,{}^a$} \\
\vspace {.1in}
\small
\begin{tabular}{c c c c c c}
\hline
\hline
\multicolumn{6}{l}{\vspace{-.12in}}\\
$~~~M~(M_{\odot})~~~$  &$R~$(km)
&$I~(10^{44}~{\rm g~cm}^2)$  & $I_u/(MR^2)$ &$~~~\kappa_n~~~$ &$~~~~n~~~~$ \\
\multicolumn{6}{l}{\vspace{-.12in}}\\
\hline
\multicolumn{6}{l}{\vspace{-.12in}}\\
0.131 &20.7 &~~0.527 &0.403 &0.117 &3.48 \\
0.154 &16.5 &~~0.595 &0.405 &0.175 &3.08 \\
0.229 &12.4 &~~0.903 &0.410 &0.314 &2.34 \\
0.344 &10.9 &~1.53 &0.417 &0.448 &1.74 \\
0.492 &10.5 &~2.51 &0.426 &0.551 &1.35 \\
0.672 &10.3 &~3.93 &0.437 &0.631 &1.07 \\
1.050 &10.4 &~7.54 &0.463 &0.726 &0.77 \\
1.470 &10.4 &12.4~~ &0.496 &0.789 &0.58 \\
1.770 &10.3 &16.4~~ &0.527 &0.830 &0.46 \\
1.930 &10.2 &18.5~~ &0.549 &0.848 &0.40 \\
2.030 &10.0 &19.6~~ &0.566 &0.857 &0.38 \\
2.110 &~~~9.69 &20.1~~ &0.588 &0.868 &0.35 \\
2.130 &~~~9.40 &19.6~~ &0.602 &0.869 &0.34 \\
2.120 &~~~9.19 &19.0~~ &0.609 &0.876 &0.33 \\
2.100 &~~~9.00 &18.2~~ &0.614 &0.876 &0.32 \\
\multicolumn{6}{l}{\vspace{-.12in}}\\
\hline
\multicolumn{6}{l}{\vspace{-.1in}}\\
\multicolumn{6}{l}{$^a$ 
Based on the AV14+UVII EOS of WFF; $I$ is the moment of inertia; $I_u$ is}\\
\multicolumn{6}{l}{the moment of inertia of a
uniform relativistic sphere with the same mass}\\
\multicolumn{6}{l}{$M$ and radius $R$ in Schwarzchild coordinates; $\kappa_n=I/I_u$.}\\
\end{tabular}
\end{center}
\end{table}

\newpage
\clearpage
\begin{table}
\begin{center}
TABLE 4\\
{\sc Binary Evolution Parameters$\,\,{}^a$} \\
\vspace {.1in}
\small
\begin{tabular}{c c c c c c c c c}
\hline
\hline
\multicolumn{8}{l}{\vspace{-.12in}}\\
 &$n$&$\bar r_m$&$\bar r_f$&$\bar v_m$&$\bar v_f$&$f_{GW}^{(max)}$(Hz)
&$\bar h_{max}$&$N_{GW}$\\
\multicolumn{9}{l}{\vspace{-.12in}}\\
\hline
\multicolumn{9}{l}{\vspace{-.12in}}\\
${\bf \hat \Omega_s=0}$ & $0.5$ & $2.76$ & $2.62$ & & & &\\
$R_o/M=5$ &     &      &      &$0.062$&$0.12$&$1302$&0.329 &$15.1$ \\
$R_o/M=8$ &     &      &      &$0.040$&$0.097$&$~~653$&0.212 &$47.6$\\
          &$1.0$&$2.49$&$2.44$& & & & \\
$R_o/M=5$ &     &      &      &$0.072$&$0.090$&$1518$&0.360 &$16.8$\\
$R_o/M=8$ &     &      &      &$0.044$&$0.063$&$~~761$&0.232 &$53.5$\\
          &$1.5$&---&$2.29$& & & & \\
$R_o/M=5$ &     &      &      &---&$0.068$&$1692$&0.375 &$18.0$\\
$R_o/M=8$ &     &      &      &---&$0.033$&$~~845$&0.240 &$57.7$\\
\multicolumn{9}{l}{\vspace{-.12in}}\\
\hline
\multicolumn{9}{l}{\vspace{-.12in}}\\
${\bf \hat \Omega_s=0.4}$ & $0.5$ & $2.67$ & $2.61$ & & & &\\
$R_o/M=5$ &     &      &      &$0.077$&$0.099$&$1387$&0.364 &$14.6$\\
$R_o/M=8$ &     &      &      &$0.051$&$0.072$&$~696$&0.235 &$45.9$\\
          &$1.0$&---&$2.48$& & & & \\
$R_o/M=5$ &     &      &      &---&$0.075$&$1515$&0.370 &$16.1$\\
$R_o/M=8$ &     &      &      &---&$0.043$&$~758$&0.237 &$51.2$\\
\multicolumn{9}{l}{\vspace{-.12in}}\\
\hline
\multicolumn{9}{l}{\vspace{-.12in}}\\
${\bf Corotation}$ & $0.5$ & $2.99$ & $2.63$ & & & &\\
$R_o/M=5$ &     &      &      &$0.066$&$0.17$&$1261$&0.308 &$10.1$\\
$R_o/M=8$ &     &      &      &$0.040$&$0.14$&$~633$&0.198 &$30.3$\\
          &$1.0$&$2.76$&$2.52$& & & &\\
$R_o/M=5$ &     &      &      &$0.074$&$0.14$&$1391$&0.323 &$12.0$\\
$R_o/M=8$ &     &      &      &$0.045$&$0.11$&$~698$&0.208 &$36.9$\\
\multicolumn{9}{l}{\vspace{-.12in}}\\
\hline
\multicolumn{9}{l}{\vspace{-.1in}}\\
\multicolumn{9}{l}{$^a$ 
Here $\bar r=r/R_o,~\bar v=v/(M/R_o)^{1/2}$; $r_m$ is the point where
$E_{eq}(r)$ is minimum; $r_f$ is the}\\
\multicolumn{9}{l}{
separation at contact; $v_m$ is the value of the radial 
velocity at $r_m$, and $v_f$ is the velocity}\\
\multicolumn{9}{l}{
at $r_f$; $f_{GW}^{(max)}=\Omega_f/\pi$ is the maximum 
frequency of the gravitational wave just prior to}\\ 
\multicolumn{9}{l}{
contact for $M=1.4M_{\odot}$, and 
$h_{max}\equiv {\bar h_{max}}M/D$ is the maximum wave amplitude for}\\
\multicolumn{9}{l}{
an observer at a distance $D$ situated along the binary axis;
$N_{GW}$ is the number of cycles}\\
\multicolumn{9}{l}{of gravitational wave from $r=5R_0$ to $r=r_f$.}\\
\end{tabular}
\end{center}
\end{table}